\documentclass[nofootinbib,twocolumn,showpacs,preprintnumbers,pre,aps,superscriptaddress]{revtex4-2}

\usepackage[dvipdfmx]{graphicx}
\usepackage{bm}
\usepackage{amsmath}
\usepackage{amssymb}
\usepackage{newtxtext}
\usepackage{newtxmath}
\usepackage{color}
\usepackage{hyperref}
\usepackage{tikz}
\usepackage{esint}
\usepackage{float} 
\allowdisplaybreaks[1]

\DeclareMathOperator{\sgn}{sgn}
\DeclareMathOperator{\erf}{erf}

\begin{document}

\title{Demon's variational principle for informational active matter}%

\author{Kento Yasuda}
\affiliation{Laboratory of Physics, College of Science and Technology, Nihon University, Funabashi, Chiba 274-8501, Japan}

\author{Kenta Ishimoto}
\affiliation{Department of Mathematics, Kyoto University, Kyoto 606-8502, Japan}

\author{Shigeyuki Komura}\email{Corresponding author: komura@wiucas.ac.cn}
\affiliation{Zhejiang Key Laboratory of Soft Matter Biomedical Materials, Wenzhou Institute, University of Chinese Academy of Sciences, Wenzhou, Zhejiang 325001, China}


\begin{abstract}
The interplay between information, dissipation, and control is reshaping our understanding of thermodynamics in 
feedback-regulated systems.
We develop the informational Onsager-Machlup principle, a generalized variational framework that unifies energetic, 
dissipative, and informational contributions within a single formalism.
This framework introduces a conditioned Onsager-Machlup integral to quantify path entropy under specified memory 
states and enables the derivation of cumulant generating functions for arbitrary observables in systems with measurement 
and feedback.
Our formulation is consistent with stochastic thermodynamics and information thermodynamics.
To demonstrate the utility of the IOMP, we apply it to an information swimmer and show that it systematically reproduces the analytical expressions for the mean velocity and higher-order cumulants obtained by a direct approach, while also enabling the calculation of entropy production in the single-measurement case.
For repeated measurements and the steady state, we derive approximate analytical expressions by using a Gaussian closure for the distribution of measured velocities. 
Our analytical results show good agreement with numerical results, except for cases of extreme drag asymmetry.
\end{abstract}

\maketitle

\section{Introduction}
\label{Intro}

Understanding the thermodynamics of systems under measurement and feedback control has become a central challenge in modern studies 
of non-equilibrium systems.
Information thermodynamics extends the traditional second law by incorporating the role of information, providing a unified framework to describe the energetics of processes involving measurement, feedback, and control~\cite{Sagawa09,Maruyama09,Parrondo15,Ito15}. 
This framework elegantly resolves longstanding conceptual puzzles such as Maxwell's demon~\cite{Strasberg13,Shiraishi15,Mandal12,Mandal13,Koski14}.
A paradigmatic example is the Szilard engine~\cite{Szilard64}, which illustrates how measurement-based feedback enables the extraction of work 
from thermal fluctuations.

While the Szilard engine idealizes Hamiltonian dynamics~\cite{Goldstein}, most realistic systems in information thermodynamics operate in 
dissipative environments.
Experimental validations using colloidal particles in viscous fluids~\cite{Toyabe10,Bauer12,Berut12,Jun14} have established a direct link between microscopic stochasticity and macroscopic energetics. 
In biological contexts, processes such as chemotaxis in microorganisms~\cite{Berg72,Wadhams04,Celani10} provide typical 
examples of information-driven regulation.

\begin{figure}[t]
\begin{center}
\includegraphics[scale=0.19]{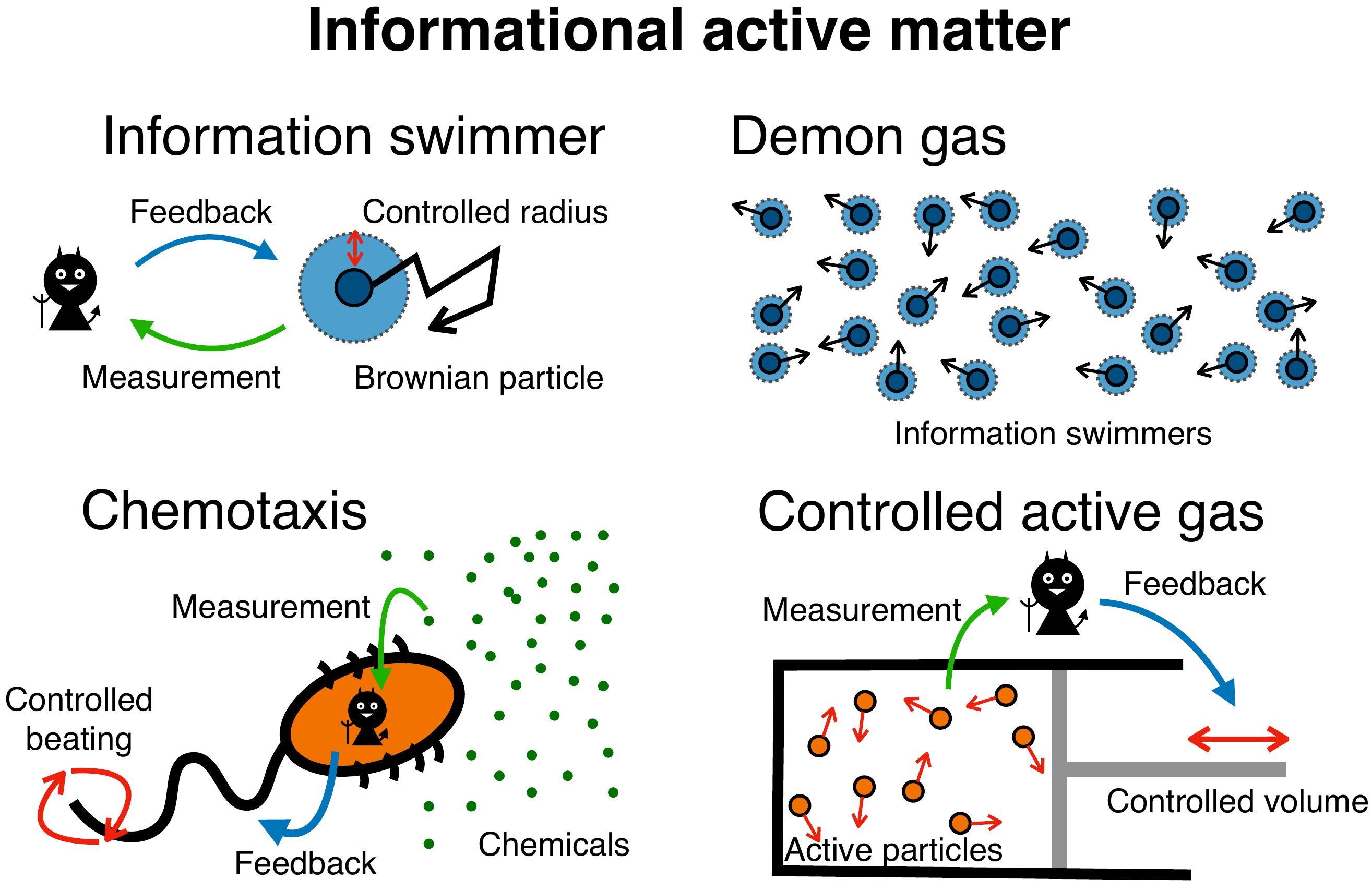}
\end{center}
\caption{
Examples of informational active matter.  
Information-fueled particle models, such as the information swimmer~\cite{Huang20,Hou24} (top left), exhibit directional motion by rectifying random Brownian motion through measurement and feedback. Beyond single-particle behavior, ensembles of information swimmers (top right) have been investigated and shown to display characteristic pattern formation~\cite{VanSaders23}.  
Particles that exploit both information and intrinsic activity~\cite{Ito15}, such as bacteria performing chemotaxis~\cite{Berg72,Wadhams04,Celani10} (bottom left), also represent important examples of 
informational active matter. 
Externally controlled active particles can also function as informational active engines~\cite{Paneru22,Saha23,Malgaretti22,Cocconi23,Cocconi24,Rafeek24,Schuttler25} (bottom right). 
}
\label{Fig:InfoActive}
\end{figure}

Building on these foundations, recent studies have investigated information engines operating in dissipative systems~\cite{Paneru20,Paneru22,Saha23}. 
Particular attention has focused on information engines powered by active particles, where autonomous energy injection 
drives persistent motion even in fluctuating environments~\cite{Malgaretti22,Cocconi23,Cocconi24,Rafeek24,Schuttler25}. 
These developments have led to the concept of informational active matter~\cite{VanSaders23}, a distinct class of systems in which information
plays a central role in generating motion and work, as illustrated in Fig.~\ref{Fig:InfoActive}. 
This paradigm differs fundamentally from conventional active matter frameworks, where particles are driven purely by energy consumption~\cite{Marchetti13,Bechinger16}.

A key model demonstrating these concepts is the information swimmer, proposed by Huang \textit{et al.}~\cite{Huang20,Hou24}. 
In this system, the particle's drag coefficient (or radius) is adaptively modulated based on measurements of its instantaneous velocity. 
Numerical simulations have shown that repeated cycles of measurement and feedback produce a finite average swimming velocity. 
Despite these advances, an analytical expression for the swimming velocity under this feedback protocol has not yet been reported.

Dynamical equations for dissipative processes can generally be derived from the Onsager principle (OP) by minimizing the instantaneous 
energy dissipation~\cite{Onsager31a,Onsager31b,Doi11,DoiBook}.
By casting the OP in a path-integral formulation, one obtains the Onsager-Machlup 
principle (OMP) as a global variational principle~\cite{Onsager53,Machlup53,Taniguchi07,Taniguchi08,Doi19}.
The OMP is a standard tool for long-time behavior, including steady states~\cite{Yasuda21JPSJ,Yasuda23}
and most probable paths~\cite{Yasuda22,Zheng25}.
In particular, the generalized Onsager-Machlup principle (GOMP), developed 
by the present authors, extends this variational framework to calculate cumulant generating 
functions (CGFs)~\cite{Yasuda24} (see Appendix~\ref{App:SOMVP} for details).
In the GOMP, however, we did not incorporate an explicit memory variable, nor did we account for 
information-based control.
Here, we further develop the GOMP framework by introducing memory and a feedback mechanism.
This extension preserves the variational structure of GOMP and enables the analysis of stochastic dynamics 
within a unified principle.

In this work, by extending the GOMP, we propose the informational Onsager-Machlup principle (IOMP), 
which we refer to as a ``demon's variational principle''~\cite{Yasuda24}.
We introduce the conditioned 
Onsager-Machlup integral (OMI), which represents the path entropy conditioned on the memory state. 
We establish a variational framework to compute the cumulant generating function (CGF) of arbitrary 
observables in measurement-feedback dynamics under a bipartite setting and 
 within a saddle-point approximation corresponding to the small-noise or low-temperature limit.
We first confirm that the IOMP is consistent with stochastic thermodynamics and information 
thermodynamics~\cite{Peliti21,Shiraishi23,Seifert25}.
To demonstrate the validity of the IOMP, we employ the information swimmer
model~\cite{Huang20} as a benchmark and derive an analytical expression for the 
average swimming velocity, in agreement with the direct calculation presented in Appendix~\ref{App:ER}.
Our analytical results show good agreement with previous numerical simulations~\cite{Huang20} 
except in the regime of extremely large drag asymmetry.
We also calculate the entropy production in the single-measurement case and clarify its dependence on the memory state.
The proposed OMI provides a variational framework for analyzing measurement-feedback dynamics and may 
serve as a useful step toward a broader theoretical description of informational active matter.

In Sec.~\ref{IOMVP}, we introduce the theoretical framework of the IOMP.
In Sec.~\ref{Information}, we discuss the thermodynamic consistency of the IOMP.
In Sec.~\ref{Model}, the information swimmer model is reviewed, and the corresponding conditioned 
OMI is derived.
Using the IOMP, we compute the CGF and the average swimming velocity for a single measurement 
and for multiple measurements in Secs.~\ref{Sec:single} and \ref{Sec:Multi}, respectively.
Finally, Sec.~\ref{Dis} summarizes our results and discusses possible applications of the IOMP.

\begin{figure}[t]
\begin{center}
\includegraphics[scale=0.24]{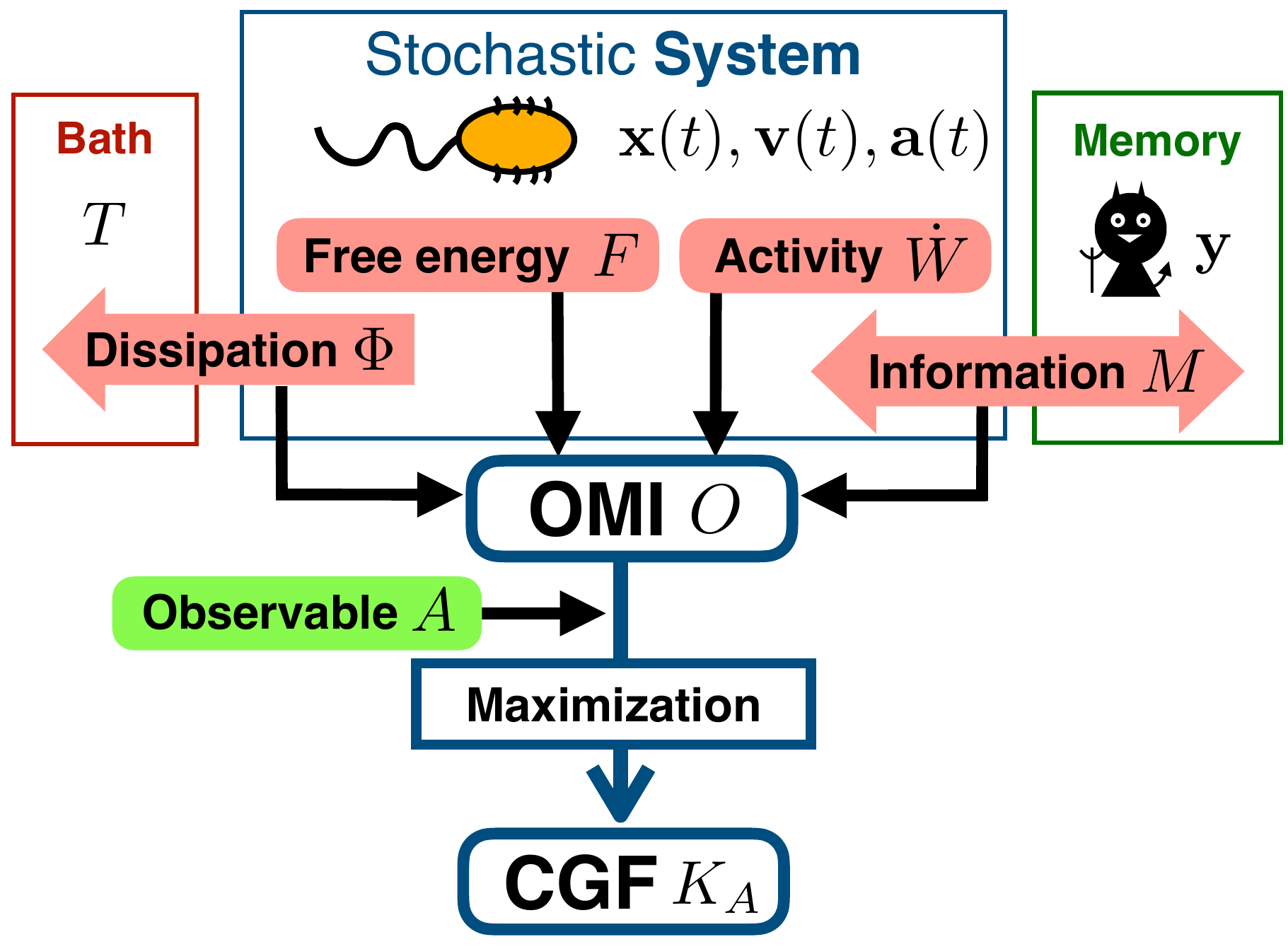}
\end{center}
\caption{
Schematic flowchart of the informational Onsager-Machlup principle (IOMP) 
[see Eqs.~(\ref{Eq:IOMVP0})-(\ref{Eq:MOM})]. 
The stochastic system, subjected to thermal fluctuations, is characterized by the free energy $F$. 
If the system is out of equilibrium, the active power $\dot{W}$ can also be incorporated.
The system is in contact with a heat bath at temperature $T$, and its energy dissipation is quantified 
by the dissipation function $\Phi$. 
Importantly, the system interacts with a memory (demon), and it is quantified by the amount 
of information, represented here by the mutual Onsager-Machlup integral (OMI) $M$ 
[see Eqs.~(\ref{mutualOMItext}) and (\ref{Eq:Info-Venn})]. 
These four quantities, free energy $F$, active power $\dot{W}$, dissipation $\Phi$, and information $M$,  
are systematically incorporated into the conditioned OMI $O$. 
For a chosen observable $A$, maximizing the modified OMI with respect to the trajectory variables 
yields the cumulant generating function (CGF) $K_A$.
}
\label{Fig:Demon}
\end{figure}

\section{Informational Onsager-Machlup principle (IOMP)}
\label{IOMVP}

We consider a system described by a state variable $\mathbf{x}$, such as the position of a colloidal particle, its 
velocity (change rate) $\mathbf{v}$, and its acceleration $\mathbf{a}$. 
In addition, we explicitly model a memory component, represented by a state variable $\mathbf{y}$, which stores the 
outcomes of measurements.
The memory state is determined by the instantaneous system state
$\mathbf{x}_0=\mathbf{x}(t_0)$ and $\mathbf{v}_0=\mathbf{v}(t_0)$
at the initial time $t=t_0$, i.e.,
$\mathbf{y}=\mathbf{y}(\mathbf{x}_0,\mathbf{v}_0)$,
and remains unchanged until the next measurement at
$t=t_{\rm f}=t_0+\tau$.
During this interval, the system evolves in contact with a thermal bath
at temperature $T$.
We are interested in a stochastic observable $A$, which depends on $\mathbf{x}$, $\mathbf{v}$, and $\mathbf{a}$.
Examples of $A$ include the velocity and position of a moving particle, as discussed later.
In general, the observable can also take the form of a vector representing multiple observables.

In our theoretical framework, we compute the cumulant generating function (CGF) of the observable $A$, defined by
$K_{A}(q) = \ln \langle \exp (qA) \rangle$, where the angular brackets denote the statistical average:
$\langle \bullet \rangle = \int d\mathbf{y} \int \mathcal{D} \mathbf{x} \, \mathcal{D} \mathbf{v} \, \mathcal{D} \mathbf{a}\,\bullet \, 
P[\mathbf{x}(t), \mathbf{v}(t), \mathbf{a}(t), \mathbf{y}]$.
Here $P[\mathbf{x}(t), \mathbf{v}(t), \mathbf{a}(t),\mathbf{y}]$ denotes the path probability distribution 
including a time-independent memory state, 
and $\int \mathcal{D} \mathbf{x} \, \mathcal{D} \mathbf{v} \, \mathcal{D} \mathbf{a}$ 
represents the path integral over all trajectories $\mathbf{x}(t)$, $\mathbf{v}(t)$, and $\mathbf{a}(t)$.
The CGF encodes the full statistical properties of $A$, and its $n$-th cumulant is obtained by 
$\langle A^n\rangle_\mathrm c=\left.d^nK_A(q)/dq^n \right|_{q=0}$~\cite{Kardarbook}.

We claim that the CGF of an informational system can be obtained by maximizing a quantity, 
referred to as the modified OMI, $\Omega_A$ [see later Eq.~(\ref{Eq:MOM})], with respect to the system 
variables $\mathbf{x}(t)$, $\mathbf{v}(t)$, and 
$\mathbf{a}(t)$ under a given memory state $\mathbf{y}$. 
We call this variational principle the informational Onsager-Machlup principle (IOMP),
which is formulated as
\begin{align}
K_A(q) = &\ln \int d \mathbf x_0d \mathbf v_0 \int d\mathbf{y} \, 
\exp \left[\Omega_A^\ast(q,\mathbf{y},\mathbf x_0,\mathbf v_0)\right]\nonumber\\
&\times p(\mathbf{y} \vert \mathbf{x}_0,\mathbf{v}_0) p(\mathbf x_0,\mathbf{v}_0),
\label{Eq:IOMVP0}
\end{align}
\begin{align}
& \Omega_A^\ast(q,\mathbf{y},\mathbf x_0,\mathbf v_0) = \max_{\mathbf x, \mathbf v,\mathbf a; \mathbf y, \mathbf x_0, \mathbf v_0} 
\Omega_A[q, \mathbf x(t), \mathbf v(t), \mathbf a(t) \vert \mathbf y, \mathbf x_0, \mathbf v_0],
\label{Eq:IOMVP}
\end{align}
where $\Omega_A[q, \mathbf x(t), \mathbf v(t), \mathbf a(t) \vert \mathbf y, \mathbf x_0, \mathbf v_0]$ is the modified OMI as we describe below.
In Eq.~(\ref{Eq:IOMVP}), the modified OMI is maximized with respect to $\mathbf{x}(t)$, $\mathbf{v}(t)$, and  $\mathbf{a}(t)$ for given 
memory state $\mathbf{y}$ and initial state $(\mathbf{x}_0, \mathbf v_0)$. 
To obtain the CGF, $\exp[\Omega_A^\ast(q,\mathbf{y},\mathbf x_0,\mathbf v_0)]$ is integrated over $\mathbf{y}$, $\mathbf x_0$, and $\mathbf v_0$,
as in Eq.~(\ref{Eq:IOMVP0}).
Moreover, $p(\mathbf{y} | \mathbf{x}_0, \mathbf v_0)$ is the conditional probability distribution for the memory state $\mathbf{y}$ under the initial state $(\mathbf{x}_0, \mathbf v_0)$, and $p(\mathbf x_0, \mathbf v_0)$ is the probability distribution for the initial state $(\mathbf{x}_0, \mathbf v_0)$.

To introduce the modified OMI, we first identify the dissipation function $\Phi(\mathbf{x}, \mathbf{v} \vert \mathbf{y})$ 
and the free energy $F(\mathbf{x},\mathbf{v} \vert \mathbf{y})$ of the system under a fixed memory state $\mathbf{y}$.
The dissipation function $\Phi$ is expressed as a quadratic form in $\mathbf{v}$ and represents the energy dissipated from the system in the noiseless limit~\cite{Onsager31a,Onsager31b,Doi11,DoiBook}.
In general, one can also consider the active power 
$\dot W(\mathbf{x}, \mathbf{v}\vert \mathbf{y})=\mathbf f(\mathbf{x}\vert \mathbf{y})\cdot \mathbf v$ 
due to the conditioned active force $\mathbf f(\mathbf x \vert \mathbf{y} )$~\cite{Wang21}, 
as well as various constraints $C$. 
By introducing $\dot W$, we can describe active Brownian particles~\cite{Zheng25} 
and other active models discussed in Ref.~\cite{Wang21}.

Using the above quantities, we define the Rayleighian as 
\begin{align}
R(\mathbf{x}, \mathbf{v}, \mathbf a\vert \mathbf{y}) & = \Phi(\mathbf{x}, \mathbf{v}\vert \mathbf{y}) 
+ \dot{F}(\mathbf{x}, \mathbf{v}, \mathbf{a}\vert \mathbf{y}) 
\nonumber \\
& \quad - \dot{W}(\mathbf{x}, \mathbf{v}\vert \mathbf{y}) + C(\mathbf{x}, \mathbf{v}, \mathbf a\vert \mathbf{y}),
\label{Rayleighian}
\end{align}
where the dot denotes a time derivative, such as $\dot{F} = dF/dt$.
In the OP, the Rayleighian $R$ is minimized with respect to $\mathbf{v}$ for 
prescribed $\mathbf{x}$ and $\mathbf{a}$ to derive the equations of motion for dissipative processes~\cite{Onsager31a,Onsager31b,Doi11,DoiBook}.
The advantage of the OP is that the obtained dynamical equations automatically satisfy Onsager's reciprocal relations 
and the second law of thermodynamics.
The OP has been successfully applied to polymers, colloidal suspensions, and active systems~\cite{Xu15,Okamoto16,Man17,Zhou18,Zhang20,Doi21}.

Next, we consider the time dependencies of $\mathbf{x}(t)$, $\mathbf{v}(t)$, $\mathbf{a}(t)$, and introduce the 
time-integrated conditioned Rayleighian, which we refer to as the conditioned 
OMI~\cite{Onsager53,Machlup53,Taniguchi07,Taniguchi08,Doi19}:
\begin{align}
&O[\mathbf x(t),\mathbf v(t),\mathbf a(t)|\mathbf y, \mathbf x_0, \mathbf v_0] \nonumber \\
 &= \frac{1}{2k_{\rm B}T}
\int_{t_0}^{t_{\rm f}} dt \, 
[R(\mathbf x(t),\mathbf v(t),\mathbf a(t)|\mathbf y) 
-R_\ast(\mathbf x(t),\mathbf a(t)|\mathbf y)].
\label{Eq:OM}
\end{align}
In the above, $k_\mathrm{B}$ is the Boltzmann constant,
$R(\mathbf x,\mathbf v,\mathbf a|\mathbf y)$ is the conditioned Rayleighian under the memory state 
$\mathbf y$, 
and $R_\ast$ represents the minimum of $R$ with respect to $\mathbf{v}$, i.e., 
$R_\ast(\mathbf x,\mathbf a|\mathbf y) = \min_{\mathbf{v}; \mathbf{x},\mathbf{a},\mathbf{y}} 
R(\mathbf x,\mathbf v,\mathbf a|\mathbf y)$.
The conditioned path probability distribution corresponding to Eq.~(\ref{Eq:OM}) is given 
by~\cite{Onsager53,Machlup53,Taniguchi07,Taniguchi08,Doi19} 
\begin{align}
&P[\mathbf{x}(t), \mathbf{v}(t),\mathbf{a}(t) \vert \mathbf{y}, \mathbf{x}_0, \mathbf v_0] \nonumber\\
&= \mathcal{N}(\mathbf{y}) 
\exp\left( -O[\mathbf{x}(t), \mathbf{v}(t),\mathbf{a}(t) \vert \mathbf{y}, \mathbf{x}_0, \mathbf v_0] \right),
\label{condOMItext}
\end{align}
where $\mathcal{N}(\mathbf{y})$ is the normalization factor.

To calculate the CGF for stochastic trajectories, we bias the conditioned OMI by 
the observable $A$ and define the modified OMI in Eq.~(\ref{Eq:IOMVP}) as~\cite{Yasuda24}
\begin{align}
\Omega_A[q, \mathbf x(t), \mathbf v(t), \mathbf a(t)|\mathbf y, \mathbf x_0, \mathbf v_0]& =qA +\ln \mathcal{N}(\mathbf{y})+\Gamma
\nonumber \\
& \quad-O[\mathbf x(t),\mathbf v(t), \mathbf a(t)|\mathbf y, \mathbf x_0, \mathbf v_0] ,
\label{Eq:MOM}
\end{align}
where $\Gamma$ represents an additional constraint, enforcing a trivial relation between $\mathbf{x}(t)$, $\mathbf{v}(t)$, 
and $\mathbf{a}(t)$, such as $\dot{\mathbf{x}} = \mathbf{v}$ and $\dot{\mathbf{v}} = \mathbf{a}$, by using a Lagrange 
multiplier (as we show later).
Equations~(\ref{Eq:IOMVP0})-(\ref{Eq:MOM}) constitute the framework of the IOMP (for detailed
derivation, see Appendix~\ref{App:SOMVP} and Ref.~\cite{Yasuda24} for the GOMP and Appendix~\ref{App:IOMVP} 
for the IOMP).
Figure~\ref{Fig:Demon} provides a schematic overview of the IOMP.

It is useful to comment on the following three points.
(i) Following standard setups in information thermodynamics~\cite{Parrondo15,Hartich14,Horowitz14}, we have assumed 
a bipartite setting in which the memory and the system do not evolve simultaneously.
Accordingly, we take $\mathbf{y}$ to be time independent.
(ii) We have employed a saddle-point approximation in Eq.~(\ref{Eq:IOMVP}), which becomes exact in the 
vanishing noise limit.
However, when applied to a functional biased by an observable as in Eq.~(\ref{Eq:MOM}), this approximation 
provides a practical route for computing fluctuating physical quantities~\cite{Touchette09,Jack20,Fodor22}. 
This is a generic approach based on the large-deviation principle, and it has been applied to a wide variety of 
problems in statistical physics. 
(iii) Although we apply the saddle-point approximation in trajectory space,
the integration over the initial state, weighted by
$p(\mathbf{x}_0,\mathbf{v}_0)$ in Eq.~(\ref{Eq:IOMVP0}), is retained.
This hybrid treatment is necessary when the memory has discrete states,
such as the binary memory of the information swimmer discussed later.
Conversely, for a continuous memory state, one could formally incorporate
the initial distribution into the action and perform a full saddle-point
evaluation, provided that the noise-amplitude scaling is uniform.

\section{Thermodynamics of IOMP}
\label{Information}

Before we apply the IOMP to an information swimmer, we mention the consistency of the 
IOMP with stochastic thermodynamics~\cite{Peliti21,Shiraishi23,Seifert25,Sekimoto11}.
We consider a trajectory $\mathbf x(t)$ evolving under a fixed memory state $\mathbf y$.
The irreversible energy dissipation $\Psi(\mathbf y)$ into the thermal bath and the microscopic degrees 
of freedom (irreversible heat dissipation) is determined by local detailed balance~\cite{Peliti21,Shiraishi23,Seifert25}
\begin{align}
\Psi(\mathbf y)=k_{\rm B} T\ln\frac{P[\mathbf x(t),\mathbf v(t),\mathbf a(t)|\mathbf y, \mathbf x_0, \mathbf v_0 ]}
{P[\mathbf x^\mathrm r(t),\mathbf v^\mathrm r(t),\mathbf a^\mathrm r(t)|\mathbf y, \mathbf x_\mathrm f, -\mathbf v_\mathrm f]},
\label{heatQ}
\end{align}
where $\mathbf x_\mathrm f=\mathbf x(t_\mathrm f)$ and $\mathbf v_\mathrm f=\mathbf v(t_\mathrm f)$ are the final states, 
and $\mathbf x^\mathrm r(t)=\mathbf x(t_0+t_{\rm f}-t)$ is the time-reversed path.
Under time reversal, the velocity and acceleration are antisymmetric and symmetric, respectively, so that we have 
$\mathbf{v}^{\mathrm r}(t) = - \mathbf{v}(t_0+t_{\rm f}-t)$ and 
$\mathbf{a}^{\mathrm r}(t) = \mathbf{a}(t_0+t_{\rm f}-t)$.
Notice that the reverse process is carried out using the same
stored value of $\mathbf y$, rather than by performing a new measurement of the reversed initial velocity.

Using Eq.~(\ref{condOMItext}), we obtain
\begin{align}
\Psi(\mathbf y)& =k_{\rm B}T(O[\mathbf{x}^\mathrm{r}(t), \mathbf{v}^\mathrm{r}(t),\mathbf{a}^\mathrm{r}(t) 
\vert \mathbf{y}, \mathbf{x}_\mathrm f,-\mathbf{v}_\mathrm f]\nonumber\\
& \quad -O[\mathbf{x}(t), \mathbf{v}(t),\mathbf{a}(t) \vert \mathbf{y}, \mathbf{x}_0, \mathbf{v}_0] )\\
& = -\Delta F(\mathbf y)+W(\mathbf y),
\label{sumEandW}
\end{align}
where we have introduced the conditioned free energy change 
$\Delta F(\mathbf y)=F(\mathbf x_\mathrm f,\mathbf v_\mathrm f |\mathbf y)
-F(\mathbf x_0,\mathbf v_0|\mathbf y)$
and the conditioned active work exerted on the system 
$W(\mathbf y)=\int_{t_0}^{t_\mathrm f}dt \, \dot{W}(\mathbf x(t),\mathbf v(t), \mathbf a(t)|\mathbf y)$.
In Eq.~(\ref{sumEandW}), we have also assumed 
$F(\mathbf x,\mathbf v |\mathbf y)=F(\mathbf x,-\mathbf v |\mathbf y)$. 
Notice that both the dissipation function $\Phi(\mathbf{x}, \mathbf{v} \vert \mathbf{y})$ 
and $R_\ast(\mathbf x,\mathbf{a}|\mathbf y)$ in the conditioned OMI do not contribute in Eq.~(\ref{sumEandW})
because they are invariant under time reversal.~\footnote{
In the noiseless limit ($T=0$), the dynamics is determined by the OP, i.e., $\delta R=0$.
In this case, the relation $2\Phi=-\dot F+\dot W$ holds, and we have 
$2\int_{t_0}^{t_\mathrm f}dt\, \Phi=-\Delta F+W=\Psi$. 
In the presence of thermal fluctuations, however, $\Phi$ no longer represents the actual dissipation, 
i.e., $2\int_{t_0}^{t_\mathrm f}dt\, \Phi\ne \Psi$.}

Let us define the mesoscopic entropy change by~\cite{Yasuda26}
\begin{align}
\Delta S_{\rm mes}(\mathbf y) = k_{\rm B} \ln \frac{p(\mathbf{x}_0, \mathbf{v}_0 \vert \mathbf y )}{
p(\mathbf{x}_\mathrm f, \mathbf{v}_\mathrm f \vert \mathbf y )},
\label{Smes}
\end{align}
where $p(\mathbf{x}_0, \mathbf{v}_0 \vert \mathbf{y})$ denotes the conditional probability distribution of the initial 
state $(\mathbf{x}_0, \mathbf{v}_0)$ for a given memory state $\mathbf{y}$.
Moreover, we have 
\begin{align}
& P[\mathbf{x}(t), \mathbf{v}(t), \mathbf{a}(t), \mathbf{x}_0, \mathbf{v}_0 \vert \mathbf{y}]  
\nonumber \\ 
& =  P[\mathbf{x}(t), \mathbf{v}(t), \mathbf{a}(t) \vert \mathbf{y}, \mathbf{x}_0, \mathbf{v}_0] 
p(\mathbf{x}_0,\mathbf{v}_0 \vert \mathbf{y}), 
\\
& P[\mathbf{x}^\mathrm r(t), \mathbf{v}^\mathrm r(t), \mathbf{a}^\mathrm r(t), \mathbf{x}_\mathrm f, -\mathbf{v}_\mathrm f \vert \mathbf{y}]  
\nonumber \\
& = P[\mathbf{x}^\mathrm r(t), \mathbf{v}^\mathrm r(t), \mathbf{a}^\mathrm r(t) \vert \mathbf{y}, \mathbf{x}_\mathrm f, -\mathbf{v}_\mathrm f] 
p(\mathbf{x}_\mathrm f,\mathbf{v}_\mathrm f \vert \mathbf{y}).
\end{align} 
With these conditioned path probability distributions,
the conditioned entropy production $\sigma(\mathbf y)$ follows from the detailed fluctuation 
theorem~\cite{Peliti21,Shiraishi23,Seifert25}
\begin{align}
\sigma(\mathbf y)&=k_{\rm B} \ln\frac{P[\mathbf x(t),\mathbf v(t),\mathbf a(t), \mathbf{x}_0,\mathbf{v}_0|\mathbf y]}
{P[\mathbf x^\mathrm r(t),\mathbf v^\mathrm r(t),\mathbf a^\mathrm r(t), \mathbf x_\mathrm f,-\mathbf v_\mathrm f|\mathbf y]},
\label{detailedFT}
\\
& = \frac{\Psi(\mathbf y)}{T} + \Delta S_{\rm mes}(\mathbf y) 
\\
& = \frac{W(\mathbf y)}{T} -\frac{\Delta F(\mathbf y)}{T}+ \Delta S_{\rm mes}(\mathbf y),
\label{entropyproduction}
\end{align}
where we have used Eqs.~(\ref{heatQ}), (\ref{sumEandW}), and (\ref{Smes}).

The change in free energy, $\Delta F$, accounts for variations in both the internal energy
and the microscopic entropy, such that $\Delta F=\Delta U-T\Delta S_{\rm mic}$~\cite{Yasuda26}. 
According to the first law of thermodynamics, the entropy change of the bath is given by $\Delta S_{\rm bath}=(W-\Delta U)/T$. 
By combining these relations, we obtain the entropy production in Eq.~(\ref{entropyproduction}) as 
$\sigma=\Delta S_{\rm bath}+\Delta S_{\rm mic}+\Delta S_{\rm mes}$~\cite{Yasuda26}.
Consequently, we confirm that the path probability defined in Eq.~(\ref{condOMItext}) within the IOMP framework is consistent with stochastic thermodynamics.
Equation~(\ref{detailedFT}) immediately leads to the integral fluctuation theorem
$\langle\!\langle e^{-\sigma(\mathbf y)/k_{\rm B}} \rangle\!\rangle =1$~\cite{Peliti21,Shiraishi23,Seifert25},
where the statistical average is defined by 
$\langle\!\langle \bullet \rangle\!\rangle = \int \mathcal{D} \mathbf{x} \, 
\mathcal{D} \mathbf{v} \, \mathcal{D} \mathbf{a}\,\bullet \, 
P[\mathbf{x}(t), \mathbf{v}(t), \mathbf{a}(t), \mathbf{x}_0,\mathbf{v}_0 \vert \mathbf y]$.
Using Jensen's inequality, we obtain the second law of thermodynamics: 
$\langle\!\langle \sigma (\mathbf y) \rangle\! \rangle \ge 0$.

Next, following information thermodynamics~\cite{Parrondo15}, we introduce the mutual information $I$ as
\begin{align}
I(\mathbf{x}, \mathbf{v}: \mathbf{y})
& = \ln p (\mathbf{x}, \mathbf{v} \vert \mathbf{y}) - \ln p (\mathbf{x}, \mathbf{v}),
\label{mutual-Infortext}
\end{align}
which quantifies the correlation between the system and the memory.
Here,  $p(\mathbf{x}, \mathbf{v}) = \int d\mathbf{y} \, p(\mathbf{x}, \mathbf{v} \vert \mathbf{y}) p(\mathbf{y})$ is the marginal probability distribution.
We decompose the mesoscopic entropy $S_{\rm mes}$ [see Eq.~(\ref{Smes})] into the mutual information 
and the apparent entropy as
$S_{\rm mes}(\mathbf{y}) = S_{\rm app} - k_{\rm B} I$, where
$S_{\rm app} = -k_{\rm B}\ln p(\mathbf{x},\mathbf{v})$ is the entropy obtained when the presence of the memory 
is ignored.
With these quantities, the second law of thermodynamics is rewritten as 
$\langle\!\langle \Psi (\mathbf{y}) \rangle\!\rangle/T + \langle\!\langle \Delta S_{\rm app} \rangle\!\rangle 
\ge k_{\rm B} \langle\!\langle \Delta I \rangle\!\rangle$, 
where $\Delta I=I(\mathbf x_{\rm f},\mathbf v_{\rm f}:\mathbf y )-I(\mathbf x_{0},\mathbf v_{0}:\mathbf y )$.
This indicates that apparent entropy production can be negative due to information loss 
($\Delta I < 0$)~\cite{Parrondo15}.

In addition, we introduce the marginal path probability distribution as 
\begin{align}
& P[\mathbf x(t),\mathbf v(t),\mathbf a(t)|\mathbf x_0,\mathbf v_0]
\nonumber \\
&=\int d\mathbf y \, 
P[\mathbf{x}(t), \mathbf{v}(t),\mathbf a(t) \vert \mathbf{y}, \mathbf{x}_0,\mathbf v_0] p(\mathbf{y} \vert \mathbf{x}_0,\mathbf v_0).
\end{align}
It is useful to define the following quantity
\begin{align}
M[\mathbf{x}(t), \mathbf{v}(t),\mathbf a(t): \mathbf{y} \vert \mathbf{x}_0,\mathbf v_0]
& = \ln P [\mathbf{x}(t), \mathbf{v}(t),\mathbf a(t) \vert \mathbf{y}, \mathbf{x}_0,\mathbf v_0]
\nonumber \\
& \quad - \ln P [\mathbf{x}(t), \mathbf{v}(t),\mathbf a(t) \vert \mathbf{x}_0,\mathbf v_0],
\label{mutualOMItext}
\end{align} 
which we call the mutual OMI, analogous to the mutual information, Eq.~(\ref{mutual-Infortext}).
The mutual OMI quantifies the strength of the correlation between the system and the memory at the trajectory level.
Then, the conditioned OMI in Eq.~(\ref{Eq:OM}) can be expressed as (see Appendix~\ref{App:MI})
\begin{align}
& O[\mathbf{x}(t), \mathbf{v}(t), \mathbf a(t) \vert \mathbf{y}, \mathbf{x}_0,\mathbf v_0] - \ln \mathcal{N}(\mathbf{y})   
\nonumber \\ 
& \quad = O[\mathbf{x}(t), \mathbf{v}(t),\mathbf a(t) \vert \mathbf x_0,\mathbf v_0]
- M[\mathbf{x}(t), \mathbf{v}(t),\mathbf a(t): \mathbf{y} \vert \mathbf{x}_0,\mathbf v_0],
\label{condOMI&mutualOMI}
\end{align}
where we have used the unconditioned OMI
\begin{align}
O[\mathbf{x}(t), \mathbf{v}(t),\mathbf a(t) \vert \mathbf x_0,\mathbf v_0] = - \ln P [\mathbf{x}(t), \mathbf{v}(t),\mathbf a(t) \vert \mathbf x_0,\mathbf v_0].
\end{align}

\section{IOMP for an information swimmer}
\label{Model}

\subsection{Model}

\begin{figure}[t]
\begin{center}
\includegraphics[scale=0.33]{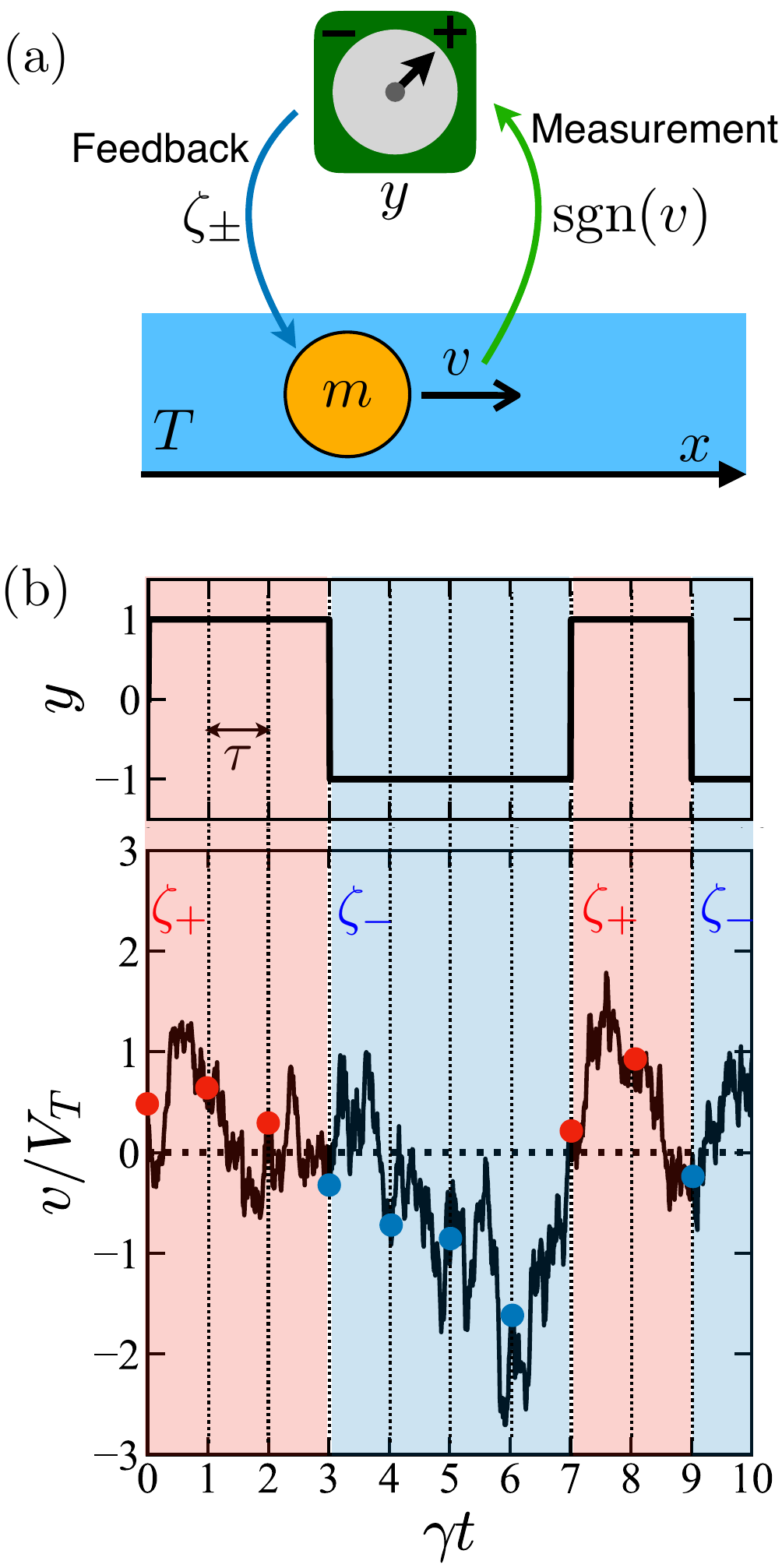}
\end{center}
\caption{
(a) Schematic illustration of the information swimmer. 
A particle of mass $m$ undergoes Brownian motion in a one-dimensional space at temperature $T$. 
At each measurement interval $\tau$, the particle velocity $v(t)$ is recorded, and its sign is stored 
in a binary memory $y = \pm 1$. 
The drag coefficient switches between $\zeta_+$ and $\zeta_-$ depending on the memory state. 
(b) Sample trajectory of the particle velocity $v$ and memory state $y$ obtained from Langevin 
simulations [see Eqs.~(\ref{Eq:Langevin}) and (\ref{Eq:FDT})], 
where the parameters are $\gamma_\pm = \zeta_\pm/m = \gamma(1 \mp \delta)$ with $\delta = 0.3$ and 
$\gamma\tau = 1$. 
Red and blue circles represent the measured velocities $V_0, V_1, V_2, \dots$, together with their signs, 
while the red and blue shaded regions indicate the corresponding memory states.
}
\label{Fig:Model}
\end{figure}

To demonstrate that the IOMP reproduces the exact analytical results, we introduce the 
information swimmer model~\cite{Huang20} as a canonical minimal model.
Before applying the IOMP, we briefly review its Langevin formulation.
As shown in Fig.~\ref{Fig:Model}(a), a particle of mass $m$ is moving with velocity $v$ in a one-dimensional dissipative 
environment at temperature $T$.
The particle experiences a viscous drag force characterized by state-dependent drag coefficients and thermal noise.
At discrete times $t = t_n$ ($n= 0, 1, \dots, N$), the particle velocity is measured and recorded as $V_n = v(t_n)$.
The measurement results are stored in memory as $y = \sgn V_n = \pm1$.
The state-dependent drag coefficient $\zeta_y$ takes different values $\zeta_{y=\pm1} = \zeta_\pm >0$ depending 
on the measurement result. 
After each feedback event, the particle undergoes random motion for a fixed time interval $\tau$, and such measurement and feedback are repeated $N$ times.
Hereafter, we refer to $\tau$ as the measurement time.

For $t_n \le t < t_{n+1}$, the Langevin equation for the particle can be written as~\cite{Huang20}   
\begin{align}
&y=\sgn V_n, \quad m\dot v=-\zeta_{y} v+\xi_n(t),
\label{Eq:Langevin}
\\
&\langle \xi_n(t)\rangle=0, \quad \langle \xi_n(t)\xi_n(t')\rangle=2k_\mathrm BT\zeta_{y}\delta(t-t'),
\label{Eq:FDT}
\end{align}
where $\xi_n(t)$ is Gaussian white noise satisfying the fluctuation-dissipation relation~\cite{DoiBook}.
The above model was investigated by numerical simulations~\cite{Huang20}, and a representative trajectory 
is shown in Fig.~\ref{Fig:Model}(b).
It was shown that the particle acquires a finite average velocity, $\langle v \rangle \ne 0$, when $\zeta_+ \ne \zeta_-$.

\subsection{Conditioned OMI}

To show the validity of the IOMP, we use the information swimmer 
model as a benchmark.
A direct derivation of the swimming velocity is presented separately in Appendix~\ref{App:ER}.
During the time interval $t_n \le t < t_{n+1}$, the dissipation function is given by $\Phi(v|y) = \zeta_y v^2/2$~\cite{DoiBook}, which is conditioned on the memory state $y=\sgn V_n$.
Since the free energy is only given by the kinetic energy of the particle, $F(v) = m v^2/2$,
the Rayleighian becomes $R(v, a |y) = \Phi(v|y) + \dot{F}(v,a) = \zeta_y v^2/2 + m a v$, where $a = \dot{v}$ is the particle 
acceleration~\cite{Machlup53,Taniguchi07,Taniguchi08}.
Minimizing $R$ with respect to $v$ while keeping $a$ fixed, we obtain 
$R_\ast(a|y) = -m^2 a^2 / (2 \zeta_y)$.
Subtracting $R_\ast$ from $R$ and performing the time integration from $t_n$ to $t_{n+1}$, 
as in Eq.~(\ref{Eq:OM}), we obtain the conditioned OMI for the information swimmer:
\begin{align}
O[v(t), a(t)| y, V_n] & = \frac{1}{2k_\mathrm BT}
\nonumber \\
& \times \int_{t_n}^{t_{n+1}} dt\, \left( \frac{\zeta_y}{2}v^2+mav+\frac{m^2}{2\zeta_y}a^2\right).
\label{Eq:COMforIS}
\end{align}

Next, we choose the observable as the particle velocity $V_{n+1}$ at time $t = t_{n+1}$.
Then the modified OMI in Eq.~(\ref{Eq:MOM}) is given by 
\begin{align}
\Omega_{V_{n+1}}[q, v(t),a(t)| y, V_n] & =qV_{n+1}-O[v(t),a(t)| y, V_n]
\nonumber \\
& \quad +\ln \mathcal{N}(y)+\Gamma,
\end{align}
where $\Gamma=\int_{t_n}^{t_{n+1}} dt\,H(t)[a(t)-\dot v(t)]$ with $H(t)$ being a Lagrange multiplier ensuring 
the relation $a=\dot{v}$.

Then, we maximize $\Omega_{V_{n+1}}[q, v(t),a(t)| y, V_n]$ with respect to $v(t)$, $a(t)$, and $H(t)$, 
under the given memory state $y$ and the initial state $V_n$. 
This is done by solving the corresponding Euler-Lagrange equations, $\delta \Omega_{V_{n+1}} = 0$, 
as shown in Appendix~\ref{AppEL}. 
By substituting the solutions of the Euler-Lagrange equations into the modified OMI, we obtain the maximized 
OMI, $\Omega_{V_{n+1}}^\ast(q, y,V_n)$.   
After fixing the normalization factor $\mathcal{N}$, it becomes 
\begin{align}
\Omega_{V_{n+1}}^\ast(q, y, V_n)
= \frac{q^2 k_\mathrm BT}{m}e^{-\gamma_y\tau}\sinh(\gamma_y \tau)
+qV_ne^{-\gamma_y \tau},
\label{Eq:obtainedJx}
\end{align}
where $\gamma_y = \zeta_y / m$ is the state-dependent decay rate.

Finally, we insert $\Omega_{V_{n+1}}^\ast$ into Eq.~(\ref{Eq:IOMVP0}) and carry out the integral over $y$. 
The integral over $y$ can be easily performed because 
the probability distribution of the memory state is simply $p(y|V_n) = \delta (y - \sgn V_n)$.
As a result, $y$ can be replaced by $\sgn V_n$ in Eq.~(\ref{Eq:obtainedJx}) 
($\gamma_y \to \gamma_{\sgn V_n}$) and 
$\Omega_{V_{n+1}}^\ast(q, y, V_n)$ becomes $\Omega_{V_{n+1}}^\ast(q, V_n)$.
In the following sections, we perform the integral over $V_n$ and evaluate the CGFs for the cases of 
single measurement and multiple measurements.

\section{Single measurement}
\label{Sec:single}

\begin{figure*}[t]
\begin{center}
\includegraphics[scale=0.33]{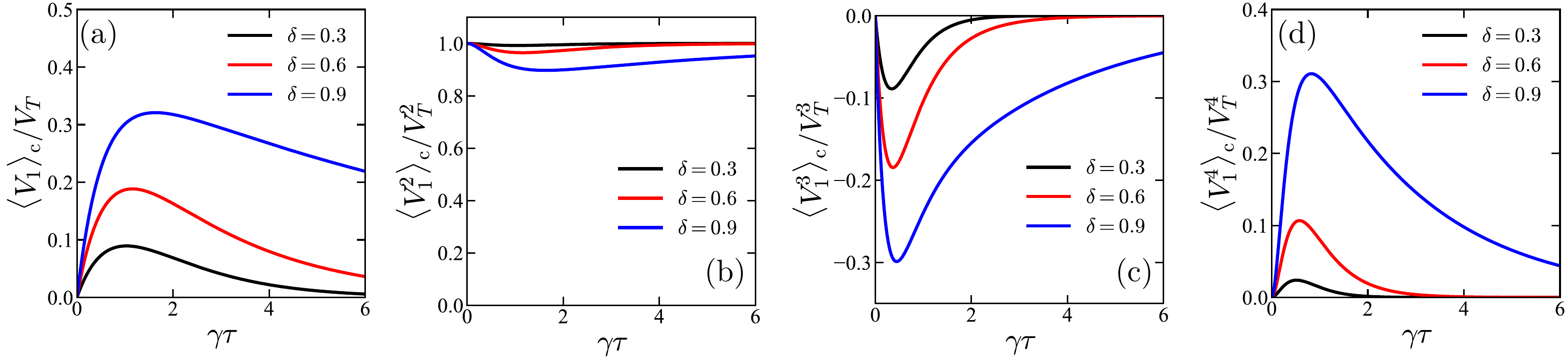}
\end{center}
\caption{
(a) First cumulant [see Eq.~(\ref{Eq:Result-V1})], (b) second cumulant [see Eq.~(\ref{Eq:Result-V1second})], 
(c) third cumulant [see Eq.~(\ref{Eq:Result-VVV1})], and (d) fourth cumulant [see Eq.~(\ref{Eq:Result-VVVV1})] of $V_1$ for 
the single measurement case as functions of the dimensionless measurement time $\gamma \tau$.
Here, $\gamma$ is defined by $\gamma_\pm = \gamma(1 \mp \delta)$, and the drag asymmetry is varied as 
$\delta=0.3$ (black), $0.6$ (red), and $0.9$ (blue).
The thermal velocity $V_T= \sqrt{k_{\rm B}T/m}$ is used to scale the velocity. 
}
\label{Fig:V-result}
\end{figure*}

\subsection{Cumulants}

We first consider the case where the measurement is performed only once at 
$t=0$ for $V_0$, and then investigate the statistical properties of the observable $V_1$ at $t=t_1=\tau$.
We assume that the probability distribution of the initial velocity $V_0$ obeys the equilibrium Maxwell-Boltzmann distribution: 
$p(V_0)=(\sqrt{2\pi}V_T)^{-1} \exp [-V_0^2/(2V_T^2)]$, where $V_T = \sqrt{k_{\rm B}T/m}$ is the thermal velocity.
From Eq.~(\ref{Eq:IOMVP0}), the CGF is obtained by the integral 
$K_{V_1}(q)=\ln \int_{-\infty}^\infty dV_0 \, \exp [\Omega_{V_1}^{\ast}(q,V_0)]  p(V_0)$. 
Since the decay rate $\gamma_\pm = \zeta_\pm / m$ is determined by $\sgn V_0$, 
the above integral can be separated into positive and negative $V_0$-regions.

As shown in Appendix~\ref{AppSingle}, the CGF for $V_1$ can be  calculated analytically as
\begin{align}
K_{V_1}(q)&=\frac{q^2 V_T^2}{2}
+\ln\left[1+\frac{1}{2} \erf \left( \frac{q V_T}{\sqrt 2}  e^{-\gamma_+\tau} \right) \right.
\nonumber \\
&\quad \left.- \frac{1}{2}\erf \left( \frac{q V_T}{\sqrt 2} e^{-\gamma_-\tau} \right) \right],
\label{Eq:Result-K}
\end{align}
where we have introduced the error function $\erf (x)=(2/\sqrt{\pi})\int_0^x dz \, e^{-z^2}$.

Each cumulant can be obtained by taking derivatives of the CGF. 
Specifically, the first cumulant is given by  
\begin{align}
\frac{\langle V_1\rangle_\mathrm c}{V_T}=\frac{1}{\sqrt{2\pi}}(e^{-\gamma_+\tau}-e^{-\gamma_-\tau}),
\label{Eq:Result-V1}
\end{align}
which corresponds to the average velocity at $t=t_1=\tau$.
This result shows that feedback through the drag coefficients leads to a finite swimming 
velocity $\langle V_1 \rangle_\mathrm{c} \neq 0$ when $\gamma_+ \neq \gamma_-$,
and it coincides with that obtained from the direct derivation in Eq.~(\ref{App:Moments}).
Moreover, Eq.~(\ref{Eq:Result-V1}) implies that $\langle V_1 \rangle_\mathrm{c} > 0$ when $\gamma_+ < \gamma_-$
and vice versa.
We further note that the first cumulant is bounded as $\langle V_1 \rangle_\mathrm{c} \le V_T/\sqrt{2\pi}$.

In Fig.~\ref{Fig:V-result}(a), we plot the first cumulant $\langle V_1 \rangle_\mathrm{c}$ as a function of the 
dimensionless measurement time $\gamma \tau$ for different values of $\delta$,
where $\gamma$ and $\delta$ are defined by $\gamma_\pm = \gamma(1 \mp \delta)$.
For small $\gamma \tau$, $\langle V_1 \rangle_\mathrm{c}$ increases linearly with $\gamma \tau$.
The average velocity reaches its maximum around $\gamma \tau \approx 1$ and subsequently decreases 
exponentially for larger $\gamma \tau$. 
This indicates that the optimal measurement time for maximizing the average velocity 
is comparable to the particle relaxation time, $1/\gamma$. 
Furthermore, the maximum value of $\langle V_1 \rangle_\mathrm{c}$ increases monotonically 
with larger $\delta$.

Similarly, the second cumulant (or the variance) can be obtained as 
\begin{align}
\frac{\langle V_1^2\rangle_\mathrm c}{V_T^2}= 
1- \frac{1}{2\pi} (e^{-\gamma_+\tau}-e^{-\gamma_-\tau})^2.
\label{Eq:Result-V1second}
\end{align}
Since $\langle V_1^2 \rangle_\mathrm{c} = \langle V_1^2 \rangle - \langle V_1 \rangle_\mathrm{c}^2$, 
where $\langle V_1^2 \rangle$ is the second moment~\cite{Kardarbook}, we obtain 
$\langle V_1^2 \rangle = V_T^2$, recovering the equipartition theorem, which can also be obtained directly from Eq.~(\ref{App:Moments}).
The third and fourth cumulants of $V_1$ are given in Eqs.~(\ref{Eq:Result-VVV1}) and~(\ref{Eq:Result-VVVV1}), 
respectively, in Appendix~\ref{AppSingle}. 
Importantly, the existence of finite higher-order cumulants implies that the probability distribution of $V_1$ 
deviates from Gaussianity when $\gamma_+ \neq \gamma_-$.

The second, third, and fourth cumulants are plotted in Fig.~\ref{Fig:V-result}(b), (c), and (d), respectively.
The third and fourth cumulants also decay exponentially with $\gamma\tau$ after exhibiting an extremum 
around $\gamma\tau \approx 1$.
These higher-order cumulants remain smaller than unity, except in the regime of large $\delta$ and small 
$\gamma\tau$. 
Hence, the probability distribution of $V_1$ can be well approximated by a Gaussian distribution 
for most parameter choices. 
In addition, we find $\langle V_1 \rangle_\mathrm{c}^2 \ll \langle V_1^2 \rangle_\mathrm{c}$, indicating that the swimming 
velocity is relatively slow compared with the diffusive process.

\begin{figure}[t]
\begin{center}
\includegraphics[scale=0.45]{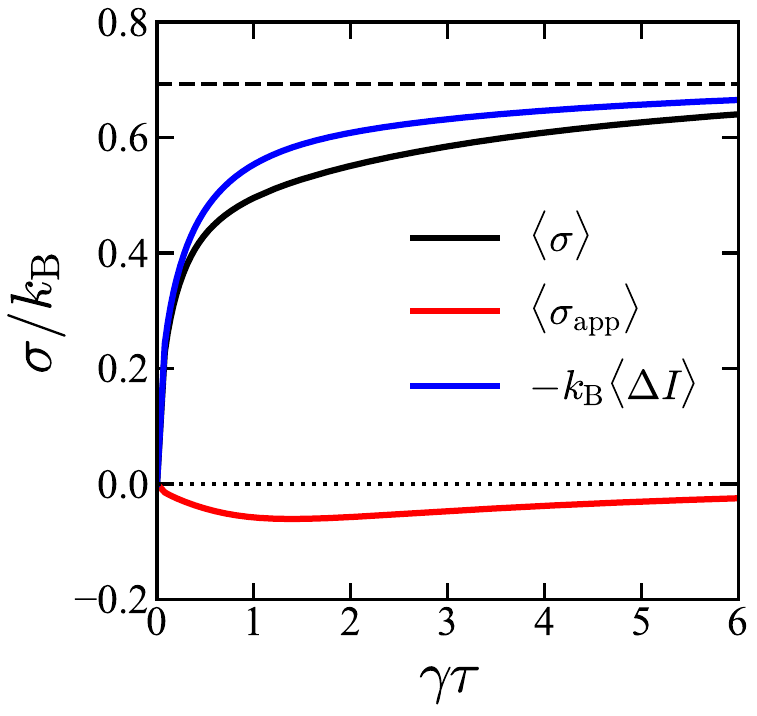}
\end{center}
\caption{
Average entropy production $\langle\sigma\rangle$ (black solid line),
apparent entropy production $\langle\sigma_{\rm app}\rangle$ (red solid line),
and information contribution $-k_{\rm B}\langle\Delta I\rangle$
(blue solid line) as functions of $\gamma\tau$ for the drag asymmetry
$\delta=0.9$.
The dashed line indicates $\sigma=k_{\rm B}\ln 2$.}
\label{Fig:Entropy2}
\end{figure}

\subsection{Entropy production}

Next, we calculate the entropy production $\sigma(y)$ [see Eq.~(\ref{entropyproduction})] 
for a specific trajectory $v(t)$ connecting $V_0\to V_1$ over a time interval of duration 
$\tau$ and for a fixed memory state $y$.
For the information swimmer, the work and free energy change are 
$W=0$ and $\Delta F=m \left(V_1^2-V_0^2 \right)/2$, respectively.
In addition, we need to calculate the mesoscopic entropy change $\Delta S_{\rm mes}$
in Eq.~(\ref{Smes}).
Using the Maxwell-Boltzmann distribution and Eq.~(\ref{App:pV1under-y}), we obtain
\begin{align}
\Delta S_{\rm mes}&=k_{\rm B}\ln \frac{p(V_0|y)}{p(V_1|y)}\\
&=\frac{m}{2T}\left(V_1^2-V_0^2 \right)-k_{\rm B}\ln\left[\frac{1}{2}+\frac{y}{2} 
\erf \left(\frac{V_1e^{-\gamma_y\tau}}{(2\langle w_y^2\rangle_\mathrm c)^{1/2}}\right)\right],
\end{align}
where $\langle w_y^2\rangle_\mathrm c=V_T^2(1-e^{-2\gamma_y \tau})$. 
Here, we have used the condition $\ln\Theta(yV_0)=0$, where $\Theta(x)$ is the Heaviside step function. 
Hence, $\sigma(y)$ becomes
\begin{align}
\sigma(y)&=-k_{\rm B}\ln\left[\frac{1}{2}+\frac{y}{2} 
\erf \left(\frac{V_1e^{-\gamma_y\tau}}{(2\langle w_y^2\rangle_\mathrm c)^{1/2}}\right)\right],
\label{Eq:SwimmerEntropy}
\end{align}
which is independent of the trajectory $v(t)$ and the initial velocity $V_0$.
In the limit $\tau\to\infty$, we have $\sigma(y)=k_{\rm B}\ln 2$, indicating that the system 
and the memory become uncorrelated after complete relaxation.
The above $\sigma(y)$ satisfies the integral fluctuation theorem, $\langle\!\langle e^{-\sigma(\mathbf y)/k_{\rm B}} \rangle\!\rangle =1$, by restoring the term $\ln\Theta(yV_0)$.

Although the above $\sigma(y)$ represents the entropy production associated
with the feedback process in informational active matter, we can also evaluate
the apparent entropy and the mutual information within the framework of
standard information thermodynamics.
The apparent entropy change is given in terms of the marginal distributions by
$\Delta S_{\rm app}= -k_{\rm B}\ln p(V_1) + k_{\rm B}\ln p(V_0)$.
Here, $p(V_0)$ is given by the Maxwell--Boltzmann distribution, whereas
$p(V_1)=\sum_{y=\pm}p(V_1|y)p(y)$ [see also Eq.~(\ref{App:pV1under-y})].
The apparent entropy production for the information swimmer is then given by
$\sigma_{\rm app}=-\Delta F/T+\Delta S_{\rm app}$.

Within standard information thermodynamics, the entropy production
$\sigma(y)$ can be decomposed into the apparent entropy production and the
change in mutual information as
$\sigma(y)=\sigma_{\rm app}-k_{\rm B}\Delta I$.
Averaging over the system and memory states using
Eq.~(\ref{App:pV1under-y}), we obtain the second law of thermodynamics in the form
$\langle\sigma\rangle = \langle\sigma_{\rm app}\rangle 
-k_{\rm B}\langle\Delta I\rangle \geq 0$.
This relation shows that the apparent entropy production can be negative
due to the loss of mutual information during the feedback process.
 
In Fig.~\ref{Fig:Entropy2}, we plot the average entropy production
$\langle\sigma\rangle$, the apparent entropy production
$\langle\sigma_{\rm app}\rangle$, and the change in mutual information
$\langle\Delta I\rangle$ as functions of $\gamma\tau$.
Although $\langle\sigma_{\rm app}\rangle$ is negative for all
$\gamma\tau$, $\langle\sigma\rangle$ remains positive, in accordance
with the second law.
In the long-time limit $\tau\to\infty$, all mutual information stored
in the binary memory is lost, yielding
$\langle\Delta I\rangle\to-\ln 2$.
In this regime, we also have
$\langle\sigma_{\rm app}\rangle
\approx -k_{\rm B}\langle V_1\rangle_{\rm c}^2/V_T^2$.
Combining these asymptotic relations with
$\langle\sigma\rangle
=\langle\sigma_{\rm app}\rangle
-k_{\rm B}\langle\Delta I\rangle\geq 0$,
we obtain the thermodynamic bound
$\langle V_1\rangle_{\rm c}^2/V_T^2 \lesssim \ln 2$.

\section{Multiple measurements}
\label{Sec:Multi}

\begin{figure*}[t]
\begin{center}
\includegraphics[scale=0.33]{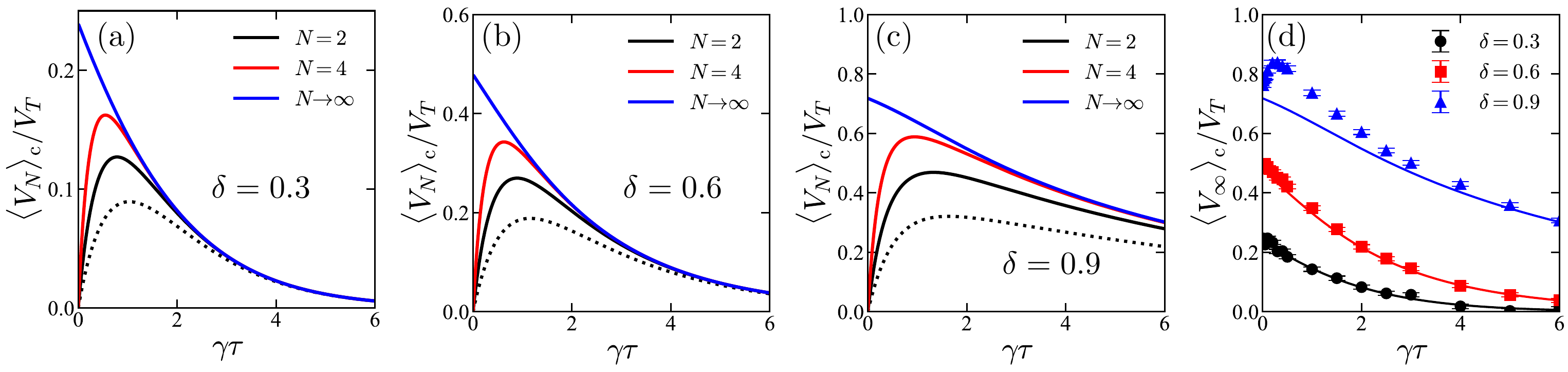}
\end{center}
\caption{
First cumulant $\langle V_N \rangle_\mathrm{c}$ for the multiple measurements case [see Eq.~(\ref{Eq:Result-VN})] as a function of the dimensionless measurement time $\gamma \tau$.
The drag asymmetry is chosen as (a) $\delta = 0.3$, (b) $\delta = 0.6$, and (c) $\delta = 0.9$. 
The black and red lines are the results for $N=2$ and $N=4$ measurements, respectively,  
while the blue lines are the steady-state velocity $\langle V_\infty \rangle_{\rm c}$ [see Eq.~(\ref{Eq:Result-Vss})].
The dotted lines are single-measurement velocity $\langle V_1 \rangle_\mathrm{c}$ [see Eq.~(\ref{Eq:Result-V1})
and Fig.~\ref{Fig:V-result}(a)].
(d) The solid lines are the steady-state velocity $\langle V_\infty \rangle_{\rm c}$ [see Eq.~(\ref{Eq:Result-Vss})] 
as functions of $\gamma \tau$ when $\delta$ is varied. 
The symbols are the results of numerical simulation of the Langevin equation in Eqs.~(\ref{Eq:Langevin}) 
and (\ref{Eq:FDT}) with $N = 2\times 10^3$ measurements. 
}
\label{Fig:Vn-result}
\end{figure*}

Next, we discuss the case of multiple measurements and the emergence of a steady-state velocity. 
As noted before, the third and fourth cumulants of $V_1$ are much smaller than 
the second cumulant for most parameter choices. 
Motivated by this observation, we employ a Gaussian closure, in which the probability distribution 
of the $n$-th measured velocity \(V_n\) is approximated by a Gaussian distribution,
$p (V_n)=(2\pi \langle V_n^2\rangle_\mathrm c)^{-1/2} 
\exp\left[-(V_n-\langle V_n\rangle_\mathrm c)^2/(2\langle V_n^2\rangle_\mathrm c)\right]$.
This approximation closes the recurrence relations at the level of the first and second cumulants.
It should be noted, however, that the exact distribution for multiple measurements need not be Gaussian, 
as already indicated by the finite third and fourth cumulants in the single-measurement case
[see Eqs.~(\ref{Eq:Result-VVV1}) and (\ref{Eq:Result-VVVV1})].

We now apply the IOMP to obtain the CGF $K_{V_{n+1}}(q)$ for $V_{n+1}$ at $t=t_{n+1}$ conditioned on  
$V_n$ at $t=t_n$. 
Estimating the maximized OMI $\Omega_{V_{n+1}}^\ast(q, V_n)$ as before, we perform the integral 
over $V_n$ with the above $p(V_n)$ (see Appendix~\ref{AppMulti}). 
To determine the cumulants, we expand $K_{V_{n+1}}(q)$ in powers of $q$ and use the approximation 
$\langle V_n \rangle_\mathrm{c}^2 \ll \langle V_n^2 \rangle_\mathrm{c}$. 
Then we obtain the following recurrence relation for the first cumulant:
\begin{align}
 \langle V_{n+1}\rangle_\mathrm c & \approx \sqrt{\frac{\langle V_n^2\rangle_\mathrm c}{2\pi}}(e^{-\gamma_+\tau}-e^{-\gamma_-\tau})
\nonumber \\
&\quad +\frac{\langle V_n\rangle_\mathrm c}{2}(e^{-\gamma_+\tau}+e^{-\gamma_-\tau}).
\label{Eq:recusion}
\end{align}
As shown in Appendix~\ref{AppMulti}, the second cumulant can be approximated as $\langle V_n^2\rangle_\mathrm c \approx V_T^2$. 
When $\langle V_0\rangle_{\rm c} = 0$, the above recurrence relation can be solved for finite measurement steps $N$:
\begin{align}
\frac{\langle V_{N}\rangle_{\mathrm{c}}}{V_T} = \sqrt{\frac{2}{\pi}} 
\frac{e^{-\gamma_{+}\tau} - e^{-\gamma_{-}\tau}}{2 - e^{-\gamma_{+}\tau} - e^{-\gamma_{-}\tau}}
\left( 1 - e^{-N/\hat{N}} \right),
\label{Eq:Result-VN}
\end{align}
where $\hat{N} = 1/\left[ \ln 2 - \ln\left( e^{-\gamma_{+}\tau} + e^{-\gamma_{-}\tau} \right) \right]$ 
is the characteristic relaxation step.

The steady-state behavior of the information swimmer is obtained by taking the limit $N \to \infty$, yielding 
\begin{align}
\frac{\langle V_\infty \rangle_{\rm c}}{V_T}= \sqrt{\frac{2}{\pi}} 
\frac{e^{-\gamma_+ \tau} - e^{-\gamma_- \tau}}{2 -  e^{-\gamma_+ \tau} - e^{-\gamma_- \tau}}.
\label{Eq:Result-Vss}
\end{align}
Unlike the single measurement case in Eq.~(\ref{Eq:Result-V1}), where $\langle V_1 \rangle_{\rm c}$ 
vanishes as $\tau \to 0$, the steady-state cumulant remains finite when $\tau \to 0$ with an expression 
$\langle V_\infty \rangle_{\rm c}/V_T \approx \sqrt{2/\pi} (\gamma_{-}-\gamma_{+})/(\gamma_{+}+\gamma_{-})$.
However, the limit $\tau \to 0$ is not physically meaningful, because $\tau$ cannot be smaller 
than the finite resolution $dt$ representing the correlation time of the discretized white noise. 
When $\gamma \tau \gg 1$, the swimmer's velocity relaxes quickly to the steady state, and 
Eqs.~(\ref{Eq:Result-V1}) and (\ref{Eq:Result-Vss}) coincide.  
When $\gamma \tau \ll 1$, on the other hand, many measurements are required to reach the steady state.

In Figs.~\ref{Fig:Vn-result}(a), (b), and (c), we plot the first cumulant $\langle V_N \rangle_\mathrm{c}$ 
in Eq.~(\ref{Eq:Result-VN}) as a function of $\gamma \tau$ for different $\delta$ values.
The black and red lines correspond to the results for $N=2$ and $4$ measurements, respectively, 
while the blue lines represent the steady-state velocity $\langle V_\infty \rangle_{\rm c}$ in Eq.~(\ref{Eq:Result-Vss}). 
The dotted lines at the bottom indicate $\langle V_1 \rangle_\mathrm{c}$ in Eq.~(\ref{Eq:Result-V1}).
In Fig.~\ref{Fig:Vn-result}(d), we plot $\langle V_\infty \rangle_{\rm c}$ as a function of $\gamma \tau$ when 
$\delta$ is varied.

To test these analytical results, we numerically solved the Langevin equation in Eqs.~(\ref{Eq:Langevin}) 
and (\ref{Eq:FDT}) using the Euler-Maruyama method, as was done in Ref.~\cite{Huang20}.
We set the initial velocity to $V_0 = 0$ and performed $N = 2\times 10^3$ measurements to reach the steady state. 
The results were averaged over $2 \times 10^4$ independent runs, and the numerical results are presented by symbols 
in Fig.~\ref{Fig:Vn-result}(d).
For $\delta = 0.3$ and $\delta = 0.6$, the analytical predictions (black and red lines) agree well with the numerical results.
For $\delta = 0.9$, however, the analytical result (blue line) deviates from the simulations at small $\gamma\tau$.

The discrepancy for $\delta=0.9$ at short measurement intervals can be
attributed to the Gaussian closure used in our analytical treatment.
For large drag asymmetry, higher-order cumulants become appreciable and
the velocity distribution deviates from a Gaussian form.
As the measurement interval increases, these non-Gaussian contributions
decrease and the agreement between the analytical prediction and the
numerical simulations improves.

\section{Summary and discussion}
\label{Dis}

In this paper, we have proposed a variational framework, termed the informational Onsager-Machlup principle (IOMP)
[see Eqs.~(\ref{Eq:IOMVP0})-(\ref{Eq:MOM})], which provides a unified approach to compute the cumulant 
generating function (CGF) for stochastic processes involving measurement and feedback.
Within the IOMP, the conditioned Onsager-Machlup integral (OMI), defined in Eq.~(\ref{Eq:OM}), plays a central 
role in quantifying the information exchange between the system and memory.
We have confirmed that the IOMP is consistent with stochastic thermodynamics.
To confirm the validity of the IOMP, we further introduced a canonical model of informational active matter,
namely, the information swimmer~\cite{Huang20}. 
By constructing the conditioned OMI [see Eq.~(\ref{Eq:COMforIS})] based on the Rayleighian under a fixed memory state, 
we evaluated the CGF of the swimming velocity. 
For a single measurement, we obtained an analytical CGF and derived the correct first and higher-order cumulants explicitly
[see Eqs.~(\ref{Eq:Result-K})-(\ref{Eq:Result-V1second}), (\ref{Eq:Result-VVV1}), and (\ref{Eq:Result-VVVV1})], 
 which coincide with those obtained from the direct derivation presented 
in Appendix~\ref{App:ER}.
We also calculated the entropy production in the single-measurement case 
[see Eq.~(\ref{Eq:SwimmerEntropy})].
For repeated measurements and the steady-state limit, we derived approximate expressions for the mean velocity
[see Eqs.~(\ref{Eq:Result-VN}) and (\ref{Eq:Result-Vss})] by employing a Gaussian closure for the distribution of
measured velocities. 
These results show that feedback through memory-dependent drag can generate a finite average velocity when the 
drag coefficients differ, $\gamma_+ \neq \gamma_-$.

The Gaussian closure also sets the range of validity of the multi-measurement and steady-state predictions. 
It neglects higher-order cumulants of $p(V_n)$, although the single measurement result already shows that the 
velocity distribution becomes non-Gaussian when $\gamma_+ \neq \gamma_-$. 
Consequently, the closure is expected to work when the third and fourth cumulants remain small, 
but it breaks down only for large drag asymmetry and short measurement intervals.
This limitation explains the deviation between Eq.~(\ref{Eq:Result-Vss}) and Langevin
simulations observed for $\delta=0.9$ at small $\gamma\tau$.

The proposed IOMP provides a unified framework for analyzing information-driven processes in nonequilibrium systems. 
Beyond the information swimmer, this approach can be extended to the design of informational engines and to the
prediction of new classes of informational active matter~\cite{VanSaders23}. 
We expect that the IOMP will provide a useful variational perspective for future studies of measurement-feedback 
control in colloidal systems, biological microswimmers, and synthetic active materials.

To demonstrate the validity of the IOMP, we have used the information swimmer model as 
a benchmark, since its results can also be obtained by direct calculation (see Appendix~\ref{App:ER}).
The IOMP provides a systematic route to obtaining the CGF not only for simple linear models but also for more complex nonlinear systems.
In particular, it can be applied to nonlinear information swimmer models for which exact results are generally unavailable.
For example, the free energy can be written as $F(x,v)=mv^2/2+U(x)$, where $U(x)$ is a nonlinear potential, such as $U(x)=\cos x$ or $U(x)=-x^2/2+x^4/4$.
The Langevin equation in Eq.~(\ref{Eq:Langevin}) can then be extended to
\begin{align}
m\dot v=-\zeta_{y} v-\partial_x U+\xi_n(t).
\end{align}
Because barrier crossing in such systems is a rare-event phenomenon, the IOMP offers 
significant advantages for its analysis.
Such applications will be discussed in future work.

In this work, we have employed the OMI, which furnishes a path-integral formulation of stochastic processes. 
Several alternative path-integral formalisms are also available, including the Feynman-Kac approach~\cite{Kac49,Wang18}, 
the Martin-Siggia-Rose-Janssen-de~Dominicis formalism~\cite{MSR73,Janssen76,Dominicis78}, and macroscopic fluctuation 
theory~\cite{Bertini15,Nardini17}.
Among these, the OMP and GOMP are particularly well suited to dissipative systems, because the dissipation functions 
entering the OMI are grounded in the Onsager principle~\cite{DoiBook}. 
We emphasize that the IOMP allows a path-integral representation that explicitly incorporates information.
However, it should be noted that the conventional OMI does not capture non-thermal noise, such as that commonly 
observed in active matter~\cite{Mizuno08,Toyota11,Ariga18,Ahmed18,Turlier16,Gnesotto18}.

So far, information thermodynamics has mainly addressed energetics and second-law-type inequalities~\cite{Parrondo15}, 
but it has not provided a systematic variational framework for deriving dynamical equations for active matter.
In this work, we have introduced an informational aspect into an energetic quantity, namely the dissipation function, 
and thereby established a method to derive the conditioned OMI that systematically represents the dynamics 
[see Eq.~(\ref{Eq:OM})].
In other words, our work provides a route to express the ideas of information thermodynamics within a path-integral formulation.

In the GOMP and IOMP, the CGF is computed by applying a saddle-point approximation at the large-deviation level
to the modified OMI [see Eqs.~(\ref{Eq:IOMVP}) and (\ref{Eq:MOM})].
Similar methods for obtaining CGFs from path integrals are well documented in the 
literature~\cite{Touchette09,Nemoto11,Nemoto14,Chetrite15,Jack20,Fodor22}.
According to the large-deviation principle, the saddle-point approximation becomes exact in the zero-noise limit. 
Moreover, by considering a biased ensemble, the saddle-point evaluation yields a CGF that encodes fluctuation information.
This is because the bias displaces the saddle point of the modified OMI from that of the original physical dynamics, 
thereby sampling fluctuations around the typical trajectory.
This approach has enabled statistical analyses of, for example, the symmetric exclusion process~\cite{Mallick22} 
and active Brownian particles~\cite{Nemoto19}.

While such path probabilities and their large-deviation structure have been widely studied, a systematic procedure for 
constructing the corresponding OMI (or action in general) is not always available and often depends on the research field.
In the macroscopic fluctuation theory, for instance, the action is constructed from the hydrodynamic 
limit~\cite{Bertini15,Nardini17}, whereas in Langevin systems, it is common to derive the action directly 
from the stochastic equation of motion~\cite{Cugliandolo17,Cugliandolo19}.
In the IOMP, we have constructed a thermodynamically consistent action from the dissipation function and the 
free energy [see Eq.~(\ref{Eq:OM})], and introduced feedback by conditioning the action on the memory.
In doing so, we have clarified a general methodology for applying large-deviation theory to dynamics with 
measurement and feedback.

During chemotaxis, microorganisms sense their environment and adjust their behavior in response to measurement 
outcomes~\cite{Berg72,Wadhams04,Celani10}.
We argue that such information-processing mechanisms in living matter can be systematically analyzed by the IOMP.
Chemotaxis can be modeled as an active Brownian particle coupled to its environment, with the persistence time of its 
rotational dynamics modulated by the measurement outcomes.
For example, the active power can be expressed as $\dot{W} = f_{\pm} v$, where the active force $f_{\pm}$ 
depends on the measurement.

The IOMP also enables us to study many-body systems of active particles that interact through information.
For example, one can consider a model in which each particle measures the velocities of its neighbors and 
adjusts its own drag coefficient accordingly.
Such information-based interactions can lead to novel phase behavior~\cite{VanSaders23}.
In this setting, one can introduce a memory-conditioned dissipation function
$\Phi\bigl(\{\mathbf{x}_i\},\{\mathbf{v}_i\}\mid\{y_i\}\bigr)
=\sum_i \zeta_i(y_i)\mathbf{v}_i^{2}/2$,
with a memory-dependent friction coefficient $\zeta_i(y_i)$ and a discrete memory variable, 
$y_i=\sgn \left( \sum_{j\in \mathrm{nbr}(i)} \mathbf{v}_i\cdot \mathbf{v}_j \right)$.
Alternatively, we can consider a scenario in which the feedback-induced change in velocity depends on the measured local particle density.
By choosing an observable such as the local orientation order parameter, one can probe the resulting phase structure.
While the optimization in Eq.~(\ref{Eq:IOMVP}) may become technically challenging for this model, the IOMP provides 
a fundamental framework for addressing the problem.

In general, the Onsager-Machlup framework can be applied to systems described by continuum fields~\cite{Yasuda24} 
representing the hydrodynamic limit of many-particle systems.
Several groups discussed the effects of non-reciprocal coupling between two phase-separating systems 
that have either non-conserved~\cite{Liu23,Tateyama24} or conserved order parameters~\cite{You20,Saha20,Suchanek23}.
It was shown that the non-reciprocality leads to the emergence of traveling patterns in purely diffusive systems, 
breaking both spatial and time-reversal symmetry.
Recently, we have found that these non-reciprocal phase separation models can be systematically derived from 
a Rayleighian with active power.
Using the Onsager-Machlup theory and its extensions, we will investigate entropy production and information flow 
to quantify the degree of non-equilibrium in non-reciprocally coupled systems~\cite{Nardini17}.

\begin{acknowledgements}

We thank Xinpeng Xu and Andrea Puglisi for useful discussions. 
K.Y.\ and K.I.\ acknowledge the Japan Society for the Promotion of Science (JSPS) KAKENHI for Transformative Research Areas A (Grant No.\ 21H05309). 
K.Y.\ acknowledges JSPS KAKENHI for Grant-in-Aid for Early-Career Scientists (Grant No.\ 25K17357). 
K.I.\ acknowledges the Japan Science and Technology Agency (JST), FOREST (Grant No.\ JPMJFR212N), and CREST (Grant No.\ JPMJCR25Q1).
S.K.\ acknowledges the support by the National Natural Science Foundation of China (Grant No.\ 12274098) and 
by the Zhejiang Key Laboratory of Soft Matter Biomedical Materials (2025ZY01036 and 2025E10072).
This work was supported by the JSPS Core-to-Core Program ``Advanced core-to-core network for the physics of self-organizing active matter" (JPJSCCA20230002).
\end{acknowledgements}

\section*{Data availability statement}
The data that support the findings of this study are available upon reasonable request from the authors.

\section*{Conflict of interest}
The authors report no conflict of interest.

\section*{Author contribution}
K.Y., K.I., and S.K.\ conceived and designed the research, contributed to writing, and reviewed and
edited the manuscript. K.Y.\ carried out the analytical calculations and prepared the figures.

\appendix
\section{Generalized Onsager-Machlup principle (GOMP)}
\label{App:SOMVP}

In this Appendix, we present the generalized Onsager-Machlup principle (GOMP), which corresponds to the memoryless 
limit of the informational Onsager-Machlup principle (IOMP) and was introduced by the present authors in 
Ref.~\cite{Yasuda24}.
For this purpose, we first review the Onsager principle (OP)~\cite{Onsager31a,Onsager31b,Doi11,DoiBook} 
and the Onsager-Machlup principle (OMP)~\cite{Onsager53,Machlup53,Doi19}. 
The OP has been employed to derive various governing equations for dissipative systems~\cite{DoiBook}.
In the absence of measurement and feedback, we focus on systems described by the state variable. 
Moreover, to account for inertial effects, we include the acceleration $\mathbf{a}$ in addition to 
$\mathbf{x}$ and $\mathbf{v}$~\cite{Machlup53,Taniguchi07,Taniguchi08}.
In the OP, we minimize the Rayleighian $R(\mathbf{x}, \mathbf{v}, \mathbf a)$ with respect to $\mathbf{v}$, 
thereby yielding the governing equations that determine the velocity $\mathbf{v}$ at a given state $\mathbf{x}$ and acceleration $\mathbf a$.
For isothermal systems, the Rayleighian is constructed as 
\begin{align}
R(\mathbf{x}, \mathbf{v}, \mathbf a) = \Phi(\mathbf{x}, \mathbf{v}) + \dot{F}(\mathbf{x}, \mathbf{v}, \mathbf{a}) - \dot{W}(\mathbf{x}, \mathbf{v}) + C(\mathbf{x}, \mathbf{v}, \mathbf a), 
\end{align}
where $\Phi(\mathbf{x}, \mathbf{v})$ is the dissipation function, $\dot{F}(\mathbf{x}, \mathbf{v}, \mathbf{a})$ is the change rate of the 
free energy, $\dot{W}(\mathbf{x}, \mathbf{v})$ is the active power, and $C(\mathbf{x}, \mathbf{v}, \mathbf a)$ denotes constraints introduced 
via Lagrange multipliers (see also Sec.~\ref{IOMVP})~\cite{DoiBook,Wang21}.

The path-integral extension of the OP is known as the OMP~\cite{Onsager53,Machlup53,Taniguchi07,Taniguchi08,Doi19}, 
which has been widely applied to various stochastic systems~\cite{Yasuda21JPSJ,Yasuda22,Yasuda23,Yasuda24}.
To formulate the OMP, we consider time-dependent variables $\mathbf{x}(t)$, $\mathbf{v}(t)$, and $\mathbf a(t)$. 
Then the Onsager-Machlup integral (OMI) is defined as~\cite{Onsager53,Machlup53,Taniguchi07,Doi19}
\begin{align}
O[\mathbf{x}(t), \mathbf{v}(t), \mathbf a(t)\vert \mathbf{x}_0,\mathbf{v}_0] & = \frac{1}{2k_{\rm B}T} \int_{t_0}^{t_{\rm f}} dt \, 
[ R(\mathbf{x}(t), \mathbf{v}(t),\mathbf a (t)) 
\nonumber \\
& \quad - R_\ast(\mathbf{x}(t),\mathbf a(t))],
\end{align}
where $(\mathbf{x}_0, \mathbf{v}_0)$ is the initial state, 
$t_0$ and $t_{\rm f}$ are the initial and final times, respectively, 
and $R_\ast(\mathbf x,\mathbf a) = \min_{\mathbf{v}; \mathbf{x}, \mathbf{a}} R(\mathbf x,\mathbf v,\mathbf a)$.

In the presence of thermal fluctuations, the system exhibits stochastic dynamics, which can be characterized by the path 
probability distribution of  $\mathbf{x}(t)$, $\mathbf{v}(t)$, and $\mathbf a(t)$.
The path probability distribution conditioned on the initial state $(\mathbf{x}_0, \mathbf{v}_0)$ is given by
\begin{align}
&P[\mathbf{x}(t), \mathbf{v}(t), \mathbf a(t) \vert \mathbf{x}_0, \mathbf{v}_0] \nonumber\\
&= \mathcal{N} \exp\left(-O[\mathbf{x}(t), \mathbf{v}(t),  \mathbf a(t)\vert \mathbf{x}_0, \mathbf{v}_0] \right),
\end{align}
where $\mathcal{N}$ is a normalization factor determined by the condition
\begin{align}
\int_{\mathbf{x}_0,\mathbf{v}_0} \mathcal{D}\mathbf{x} \, \mathcal{D}\mathbf{v} \,  \mathcal{D}\mathbf a\,P[\mathbf{x}(t), \mathbf{v}(t) ,\mathbf a(t)\vert \mathbf{x}_0, \mathbf{v}_0] = 1.
\end{align}
Here, $\int_{\mathbf{x}_0,\mathbf{v}_0} \mathcal{D}\mathbf{x} \, \mathcal{D}\mathbf{v}\, \mathcal{D}\mathbf{a}$ denotes the path integral over all trajectories satisfying 
the initial condition $(\mathbf{x}_0, \mathbf{v}_0)$.
In the OMP, the minimum of the OMI gives the equations for the most probable path~\cite{Onsager53,Machlup53,Taniguchi07,Taniguchi08,Doi19}.

Recently, we proposed the generalized Onsager-Machlup principle (GOMP), and provided a method for obtaining the 
cumulant generating function (CGF) within a variational framework~\cite{Yasuda24}. 
The GOMP enables a systematic computation of the CGF for observables, such as the position 
of a colloidal particle, without the need for explicit solutions of the underlying stochastic differential equations.
The GOMP successfully describes the fluctuating dynamics of fluids under both equilibrium and non-equilibrium conditions~\cite{Yasuda24}.

Consider the CGF of an observable $A$, defined as $K_A(q) = \ln \langle \exp(qA) \rangle$, 
where $\langle \bullet \rangle$ denotes the statistical average over trajectories, i.e., 
$\langle \bullet \rangle = \int \mathcal{D}\mathbf{x} \, \mathcal{D}\mathbf{v} \,\mathcal{D}\mathbf a\, \bullet \, P[\mathbf{x}(t), \mathbf{v}(t), \mathbf a(t)]$.
Here, the path integral is taken over all realizations of $\mathbf{x}(t)$, $\mathbf{v}(t)$, and $\mathbf a(t)$,
and the full path probability distribution $P[\mathbf{x}(t), \mathbf{v}(t), \mathbf  a(t)]$ is given by
\begin{align}
P[\mathbf{x}(t), \mathbf{v}(t),\mathbf a(t)] = P[\mathbf{x}(t), \mathbf{v}(t),\mathbf a(t) \vert \mathbf{x}_0,\mathbf{v}_0] p(\mathbf{x}_0,\mathbf{v}_0), 
\end{align}
where $p(\mathbf{x}_0,\mathbf{v}_0)$ is the probability distribution for the initial state $(\mathbf{x}_0, \mathbf{v}_0)$.
Hence, the CGF can be written as  
\begin{align}
K_A(q) & = \ln \int \mathcal{D} \mathbf{x} \, \mathcal{D} \mathbf{v} \, \mathcal{D} \mathbf{a} \, \exp(qA) P[\mathbf{x}(t), \mathbf{v}(t),\mathbf{a}(t)]
\\
& = \ln \int d\mathbf{x}_0d\mathbf{v}_0 \int_{\mathbf{x}_0, \mathbf{v}_0} \mathcal{D} \mathbf{x} \, \mathcal{D} \mathbf{v} \, \mathcal{D} \mathbf{a} \,\exp(qA) 
\nonumber \\
& \quad \times P[\mathbf{x}(t), \mathbf{v}(t),\mathbf{a}(t) \vert \mathbf{x}_0,\mathbf{v}_0] p(\mathbf{x}_0,\mathbf{v}_0).
\end{align}
With this CGF, the $n$-th cumulant is obtained by 
$\langle A^n\rangle_\mathrm c=\left.d^nK_A(q)/dq^n \right|_{q=0}$.

Applying the saddle-point approximation with respect to $\mathbf{x}(t)$, $\mathbf{v}(t)$, and $\mathbf{a}(t)$~\cite{Touchette09}, 
one can obtain the CGF through the following GOMP~\cite{Yasuda24}:
\begin{align}
& K_A(q) = \ln \int d\mathbf{x}_0d\mathbf{v}_0 \,  \exp\left[ \Omega_A^\ast(q, \mathbf{x}_0,\mathbf{v}_0) \right] p(\mathbf{x}_0,\mathbf{v}_0), 
\label{KAqapp}
\\
& \Omega_A^\ast(q, \mathbf{x}_0,\mathbf{v}_0) = \max_{\mathbf{x}, \mathbf{v}, \mathbf{a}; \, \mathbf{x}_0,\mathbf{v}_0} 
\Omega_A [q, \mathbf{x}(t), \mathbf{v}(t), \mathbf{a}(t) \vert \mathbf{x}_0,\mathbf{v}_0], 
\label{Eq:SOMVP}
\end{align}
where the modified OMI $\Omega_A[q, \mathbf{x}(t), \mathbf{v}(t), \mathbf{a}(t) \vert \mathbf{x}_0,\mathbf{v}_0]$ is introduced by
\begin{align}
\Omega_A[q, \mathbf{x}(t), \mathbf{v}(t), \mathbf{a}(t)\vert \mathbf{x}_0,\mathbf{v}_0]  
& = qA  + \ln \mathcal{N} + \Gamma
\nonumber\\
& \quad - O[\mathbf{x}(t), \mathbf{v}(t), \mathbf{a}(t) \vert \mathbf{x}_0,\mathbf{v}_0].
\label{Legendre}
\end{align}
The last term $\Gamma$ represents an additional constraint that enforces the trivial kinematic relation between $\mathbf{x}$, 
$\mathbf{v}$, and $\mathbf{a}$, typically of the form $\dot{\mathbf{x}} = \mathbf{v}$ and $\dot{\mathbf{v}} = \mathbf{a}$.
In Eq.~(\ref{KAqapp}), we take the average over $\mathbf{x}_0$ and $\mathbf{v}_0$ since the initial state is randomly sampled from the probability distribution $p(\mathbf{x}_0,\mathbf{v}_0)$.

\section{Informational Onsager-Machlup principle (IOMP)}
\label{App:IOMVP}

In this Appendix, we extend the GOMP to incorporate measurement and feedback processes, thereby formulating the 
informational Onsager-Machlup principle (IOMP). 
To this end, we introduce an additional variable $\mathbf{y}$ representing the memory state.
The Rayleighian for an isothermal system is the same as that in Appendix~\ref{App:SOMVP}.
In the context of information theory, all quantities are now defined under a fixed memory state $\mathbf{y}$, such as  
$\Phi = \Phi(\mathbf{x}, \mathbf{v} \vert \mathbf{y})$, 
$\dot{F} = \dot{F}(\mathbf{x}, \mathbf{v}, \mathbf{a} \vert \mathbf{y})$,
and $\dot W(\mathbf{x}, \mathbf{v}\vert \mathbf{y})$.

By considering the time evolution of the system variables $\mathbf{x}(t)$, $\mathbf{v}(t)$, and $\mathbf{a}(t)$, the conditioned OMI 
$O[\mathbf x(t),\mathbf v(t),\mathbf{a}(t)|\mathbf y, \mathbf{x}_0, \mathbf{v}_0]$ is given by 
\begin{align}
&O[\mathbf x(t),\mathbf v(t),\mathbf{a}(t)|\mathbf y, \mathbf x_0,\mathbf{v}_0] \nonumber\\
&= \frac{1}{2k_{\rm B}T}
\int_{t_0}^{t_{\rm f}}  dt \, 
[R(\mathbf x(t),\mathbf v(t),\mathbf{a}(t)|\mathbf y) 
 -R_\ast(\mathbf x(t),\mathbf{a}(t)|\mathbf y)],
\label{Eq:OMapp}
\end{align}
where $R_\ast(\mathbf x,\mathbf{a}|\mathbf y) = \min_{\mathbf{v}; \mathbf{x},\mathbf{a},\mathbf{y}} 
R(\mathbf x,\mathbf v,\mathbf{a}|\mathbf y)$ [see Eq.~(\ref{Eq:OM})].
The conditioned OMI represents the entropy of a certain path conditioned on a given memory state $\mathbf{y}$ and 
the initial state $(\mathbf{x}_0, \mathbf{v}_0)$.
The corresponding conditioned path probability distribution is given by 
\begin{align}
&P[\mathbf{x}(t), \mathbf{v}(t),\mathbf{a}(t) \vert \mathbf{y}, \mathbf{x}_0, \mathbf{v}_0] \nonumber\\
&= \mathcal{N}(\mathbf{y}) 
\exp\left( -O[\mathbf{x}(t), \mathbf{v}(t),\mathbf{a}(t) \vert \mathbf{y}, \mathbf{x}_0, \mathbf{v}_0] \right),
\label{condOMIapp}
\end{align}
where $\mathcal{N}(\mathbf{y})$ is a normalization factor fixed by 
\begin{align}
\int_{\mathbf{x}_0,\mathbf{v}_0} \mathcal{D}\mathbf{x} \, \mathcal{D}\mathbf{v} \,\mathcal{D}\mathbf{a}\, P[\mathbf{x}(t), \mathbf{v}(t),\mathbf{a}(t) \vert \mathbf{y}, \mathbf{x}_0,\mathbf{v}_0] = 1.
\end{align}

The joint path probability distribution $P[\mathbf{x}(t), \mathbf{v}(t), \mathbf{a}(t), \mathbf{y} \vert \mathbf{x}_0,\mathbf{v}_0]$ of the system 
and the memory conditioned on the initial state is given by
\begin{align}
&P[\mathbf{x}(t), \mathbf{v}(t), \mathbf{a}(t), \mathbf{y} \vert \mathbf{x}_0,\mathbf{v}_0]\nonumber\\
& = 
P[\mathbf{x}(t), \mathbf{v}(t), \mathbf{a}(t) \vert \mathbf{y}, \mathbf{x}_0, \mathbf{v}_0] p(\mathbf{y} \vert \mathbf{x}_0,\mathbf{v}_0),
\label{Eq:ConditionedProb}
\end{align}
where $p(\mathbf{y} \vert \mathbf{x}_0,\mathbf{v}_0)$ denotes the conditional probability distribution of the memory state 
$\mathbf{y}$ for a given initial state $(\mathbf{x}_0, \mathbf{v}_0)$.
The full joint path probability distribution $P[\mathbf{x}(t), \mathbf{v}(t),\mathbf{a}(t),  \mathbf{y}]$ without conditioning on the initial state 
is given by
\begin{align}
&P[\mathbf{x}(t), \mathbf{v}(t),\mathbf{a}(t),  \mathbf{y}] \nonumber \\
&= P[\mathbf{x}(t), \mathbf{v}(t), \mathbf{a}(t), \mathbf{y} \vert \mathbf{x}_0,\mathbf{v}_0] p(\mathbf{x}_0,\mathbf{v}_0).
\end{align}
Hence, the CGF of an observable $A$ can be written as
\begin{align}
K_A(q) & = \ln \int d\mathbf{y} \int \mathcal{D} \mathbf{x} \, \mathcal{D} \mathbf{v}\, \mathcal{D} \mathbf{a} \, \exp(qA) \nonumber \\
& \quad \times P[\mathbf{x}(t), \mathbf{v}(t), \mathbf{a}(t), \mathbf{y}]
\nonumber \\
& = \ln \int d\mathbf{x}_0d\mathbf{v}_0 \, \int d\mathbf{y} \, \int_{\mathbf{x}_0,\mathbf{v}_0} \mathcal{D} \mathbf{x} \, \mathcal{D} \mathbf{v} \,\mathcal{D} \mathbf{a} \, \exp(qA) 
\nonumber \\
& \quad \times P[\mathbf{x}(t), \mathbf{v}(t), \mathbf{a}(t) \vert \mathbf{y}, \mathbf{x}_0,\mathbf{v}_0] p(\mathbf{y} \vert \mathbf{x}_0,\mathbf{v}_0) p(\mathbf{x}_0,\mathbf{v}_0).
\end{align}

Applying the saddle-point approximation with respect to $\mathbf{x}(t)$, $\mathbf{v}(t)$, and $\mathbf{a}(t)$ as before~\cite{Yasuda24,Touchette09}, 
we obtain 
\begin{align}
 K_A(q) =& \ln \int d \mathbf x_0d\mathbf{v}_0 \int d\mathbf{y} \, 
\exp \left[\Omega_A^\ast(q,\mathbf{y},\mathbf x_0,\mathbf{v}_0)\right]\nonumber\\
&\times p(\mathbf{y} \vert \mathbf{x}_0,\mathbf{v}_0) p(\mathbf x_0,\mathbf{v}_0),
\label{Eq:IOMVP0app}
\end{align}
\begin{align}
& \Omega_A^\ast(q,\mathbf{y},\mathbf x_0,\mathbf{v}_0) = \max_{\mathbf x, \mathbf v, \mathbf a; \mathbf y, \mathbf x_0,\mathbf{v}_0} 
\Omega_A[q, \mathbf x(t), \mathbf v(t), \mathbf a(t) \vert \mathbf y, \mathbf x_0,\mathbf{v}_0],
\label{Eq:IOMVPapp}
\end{align}
[see Eqs.~(\ref{Eq:IOMVP0}) and (\ref{Eq:IOMVP})]. 
In the above, the modified OMI is introduced by 
\begin{align}
\Omega_A[q, \mathbf x(t), \mathbf v(t), \mathbf a(t)|\mathbf y, \mathbf x_0,\mathbf{v}_0] & =qA +\ln \mathcal{N}(\mathbf{y})+\Gamma
\nonumber \\
& \quad-
O[\mathbf x(t),\mathbf v(t), \mathbf a(t)|\mathbf y, \mathbf x_0,\mathbf{v}_0] ,
\label{Eq:MOMXapp}
\end{align}
[see Eq.~(\ref{Eq:MOM})].
Equations~(\ref{Eq:OMapp}), (\ref{Eq:IOMVP0app}), (\ref{Eq:IOMVPapp}), and (\ref{Eq:MOMXapp}) constitute the IOMP.

\section{Mutual Onsager-Machlup integral}
\label{App:MI}

From the joint path probability distribution $P[\mathbf{x}(t), \mathbf{v}(t), \mathbf a(t) ,\mathbf{y} \vert \mathbf{x}_0, \mathbf{v}_0]$ in Eq.~(\ref{Eq:ConditionedProb}),
one can obtain two marginal probability distributions:
\begin{align}
& P[\mathbf x(t),\mathbf v(t), \mathbf a(t)|\mathbf x_0, \mathbf{v}_0]=\int d\mathbf y \, P[\mathbf x(t),\mathbf v(t), \mathbf a(t), \mathbf y|\mathbf x_0, \mathbf{v}_0],
\\
& p(\mathbf y|\mathbf x_0, \mathbf{v}_0)=\int_{\mathbf x_0,\mathbf{v}_0} \mathcal D \mathbf x \, \mathcal D \mathbf v\, \mathcal D \mathbf a \,
P[\mathbf x(t),\mathbf v(t),\mathbf a(t),\mathbf y|\mathbf x_0,\mathbf{v}_0]. 
\end{align}
Then we introduce the following quantity, which we call the mutual OMI  
\begin{align}
M[\mathbf{x}(t), \mathbf{v}(t),\mathbf{a}(t): \mathbf{y} \vert \mathbf{x}_0,\mathbf{v}_0]
& = \ln P [\mathbf{x}(t), \mathbf{v}(t), \mathbf{a}(t) \vert \mathbf{y}, \mathbf{x}_0,\mathbf{v}_0]
\nonumber \\
& \quad - \ln P [\mathbf{x}(t), \mathbf{v}(t), \mathbf{a}(t) \vert \mathbf{x}_0,\mathbf{v}_0], 
\label{Eq:Info-Venn}
\end{align}
where $P [\mathbf{x}(t), \mathbf{v}(t), \mathbf{a}(t) \vert \mathbf{y}, \mathbf{x}_0,\mathbf{v}_0]$ is the conditioned 
path probability distribution in Eq.~(\ref{condOMIapp}).
The mutual OMI is analogous to mutual information in information thermodynamics~\cite{Parrondo15},
and quantifies the strength of the correlation between the system and the memory.

By using these probability distributions, Eq.~(\ref{Eq:ConditionedProb}) can be rewritten as 
\begin{align}
&\ln P[\mathbf x(t),\mathbf v(t), \mathbf{a}(t),\mathbf y \vert \mathbf x_0,\mathbf{v}_0]  =
\ln P [\mathbf{x}(t), \mathbf{v}(t), \mathbf{a}(t) \vert \mathbf x_0,\mathbf{v}_0] \nonumber \\
& \quad + \ln p(\mathbf {y} \vert \mathbf x_0,\mathbf{v}_0)
+ M[\mathbf{x}(t), \mathbf{v}(t), \mathbf{a}(t): \mathbf{y} \vert \mathbf{x}_0,\mathbf{v}_0].
\label{totalOMIapp}
\end{align}
If we further define the total OMI, the unconditioned OMI, and the memory OMI by  
\begin{align}
&O[\mathbf x(t),\mathbf v(t), \mathbf{a}(t),\mathbf y \vert \mathbf x_0,\mathbf{v}_0] = -\ln P[\mathbf x(t),\mathbf v(t), \mathbf{a}(t),\mathbf y \vert \mathbf x_0,\mathbf{v}_0],
\\
&O[\mathbf{x}(t), \mathbf{v}(t), \mathbf{a}(t) \vert \mathbf x_0,\mathbf{v}_0] = - \ln P [\mathbf{x}(t), \mathbf{v}(t), \mathbf{a}(t) \vert \mathbf x_0,\mathbf{v}_0], 
\\
&O(\mathbf {y} \vert \mathbf x_0,\mathbf{v}_0)=- \ln p(\mathbf {y} \vert \mathbf x_0,\mathbf{v}_0),
\label{OMIs}
\end{align}
respectively, Eq.~(\ref{totalOMIapp}) can also be expressed as~\cite{Parrondo15}  
\begin{align}
&O[\mathbf x(t),\mathbf v(t), \mathbf{a}(t),\mathbf y \vert \mathbf x_0, \mathbf{v}_0] =
O[\mathbf{x}(t), \mathbf{v}(t), \mathbf{a}(t) \vert \mathbf x_0,\mathbf{v}_0]\nonumber \\
& \quad  + O(\mathbf {y} \vert \mathbf x_0, \mathbf{v}_0)
- M[\mathbf{x}(t), \mathbf{v}(t), \mathbf{a}(t): \mathbf{y} \vert \mathbf{x}_0,\mathbf{v}_0].
\label{totalOMIsum}
\end{align}
If we use Eq.~(\ref{condOMIapp}), the conditioned OMI in Eq.~(\ref{Eq:OMapp}) can be written as 
\begin{align}
& O[\mathbf{x}(t), \mathbf{v}(t), \mathbf{a}(t) \vert \mathbf{y}, \mathbf{x}_0, \mathbf{v}_0] - \ln \mathcal{N}(\mathbf{y})   
\nonumber \\ 
& \quad = O[\mathbf{x}(t), \mathbf{v}(t), \mathbf{a}(t) \vert \mathbf x_0,\mathbf{v}_0]
- M[\mathbf{x}(t), \mathbf{v}(t), \mathbf{a}(t): \mathbf{y} \vert \mathbf{x}_0,\mathbf{v}_0].
\end{align}

Using the mutual OMI in Eq.~(\ref{Eq:Info-Venn}), we can rewrite the irreversible heat dissipation in Eq.~(\ref{heatQ}) as 
\begin{align}
\Psi(\mathbf y) &=  \Psi_{\rm app}-k_{\rm B} T \Delta M(\mathbf y),
\label{appQ}
\end{align}
where $\Psi_{\rm app}$ is the apparent dissipation when the presence of memory is ignored 
\begin{align}
\Psi_{\rm app} = k_{\rm B}T\ln\frac{P[\mathbf x(t),\mathbf v(t),\mathbf a(t)| \mathbf x_0,\mathbf{v}_0]}
{P[\mathbf x^\mathrm r(t),\mathbf v^\mathrm r(t),\mathbf a^\mathrm r(t)|\mathbf x_\mathrm f,-\mathbf{v}_\mathrm f]},
\end{align}
which is independent of $\mathbf y$, and 
\begin{align}
\Delta M (\mathbf y) & = M[\mathbf x^\mathrm r(t), \mathbf{v}^\mathrm r(t), \mathbf{a}^\mathrm r(t): 
\mathbf y|\mathbf x_\mathrm f,-\mathbf{v}_\mathrm f]
\nonumber \\
& \quad - M[\mathbf x(t), \mathbf{v}(t), \mathbf{a}(t): \mathbf y|\mathbf x_0,\mathbf{v}_0], 
\end{align}
is the difference in the mutual OMI.
Hence, $\Delta M(\mathbf y)$ provides a consistent correction to the apparent dissipation $\Psi_{\rm app}$, 
yielding $\Psi(\mathbf y)$.

\section{Direct approach for an information swimmer}
\label{App:ER}

We can obtain the exact probability distribution of the information swimmer 
because the model in Eqs.~(\ref{Eq:Langevin}) and (\ref{Eq:FDT}) is linear and the 
swimmer behaves as a free Brownian particle after the measurement. 
The conditioned probability distribution $p(V_1|V_0, y)$ can then be calculated as
\begin{align}
&p(V_1|V_0, y)=\frac{1}{(2\pi\langle w_y^2\rangle_\mathrm c)^{1/2}}
\exp\left[-\frac{(V_1-V_0 e^{-\gamma_y \tau})^2}{2\langle w_y^2\rangle_\mathrm c} \right],
\end{align}
where $\langle w_y^2\rangle_\mathrm c=V_T^2(1-e^{-2\gamma_y \tau})$.
Furthermore, we can calculate the probability distribution of $V_1$ conditioned on the memory state as 
$p(V_1|y)=\int_{-\infty}^\infty dV_0\, p(V_1|V_0, y)p(V_0|y)$, where $p(V_0|y)$ is the probability 
distribution of $V_0$ conditioned on the memory state $y$.
For the information swimmer, we have $p(V_0|y)=2p(V_0)\Theta(yV_0)$, where $p(V_0)$ is the 
Maxwell-Boltzmann distribution defined in Sec.~\ref{Sec:single} and $\Theta(x)$ is the Heaviside step function.

After some calculation, we obtain
\begin{align}
&p(V_1|y)=\frac{e^{-V_1^2/(2V_T^2)}}{(2\pi)^{1/2}V_T}\left[1+y \erf \left(\frac{V_1e^{-\gamma_y\tau}}
{(2\langle w_y^2\rangle_\mathrm c)^{1/2} }\right)\right].
\label{App:pV1under-y}
\end{align}
Using the above result, we can calculate the $n$th moment by
\begin{align}
\langle V_1^n\rangle=\int_{-\infty}^\infty dV_1\, V_1^n\sum_{y=1,-1}\frac{p(V_1|y)}{2}.
\label{App:Moments}
\end{align}
For $n=1$ and $2$, these moments coincide with the corresponding results
obtained via the IOMP in Eqs.~(\ref{Eq:Result-V1}) and (\ref{Eq:Result-V1second}), respectively.

\section{Euler-Lagrange equations for the information swimmer}
\label{AppEL}

Here, we present the Euler-Lagrange equations for the information swimmer and their solutions.  
By using Eq.~(\ref{Eq:COMforIS}) and taking the first variation of $\Omega_{V_{n+1}}[q, v(t),a(t)| y, V_n]$ with respect to 
$v(t)$, $a(t)$, and $H(t)$, and setting it to zero, we obtain the following Euler-Lagrange equations:
\begin{align}
&\zeta_y v - \frac{m^2}{\zeta_y} \ddot{v} = 0 \quad (t_n \le t \le t_{n+1}),
\label{Eq:ELE} \\
&H = \frac{1}{2k_{\mathrm{B}}T} \left( m v + \frac{m^2}{\zeta_y} a \right),
\end{align}
together with the natural final condition $q = H(t_{n+1})$.
We impose the initial condition $v(t_n) = V_n$ to solve Eq.~(\ref{Eq:ELE}) as 
$v(t) = C_1 e^{\gamma_y (t - t_n)} + C_2 e^{-\gamma_y (t - t_n)}$,
where $\gamma_y = \zeta_y / m$.  
The coefficients $C_1$ and $C_2$ are determined by the initial and final conditions as 
\begin{align}
v(t) = \frac{2k_{\mathrm{B}}T q}{m} e^{-\gamma_y \tau} \sinh\!\left[ \gamma_y (t - t_n) \right]
+ V_n e^{-\gamma_y (t - t_n)}.
\end{align}
Substituting this solution into the conditioned OMI and fixing the normalization factor $\mathcal{N}$, we 
arrive at Eq.~(\ref{Eq:obtainedJx}).

\section{Single measurement}
\label{AppSingle}

As shown in Sec.~\ref{Sec:single}, the CGF for a single measurement is given by 
\begin{align}
K_{V_1}(q) = \ln \int_{-\infty}^{\infty} dV_0 \, \exp\left[ \Omega_{V_1} ^\ast (q, V_0) \right] p(V_0),
\end{align}
where $\Omega_{V_1}^\ast  (q, V_0)$ is given by Eq.~(\ref{Eq:obtainedJx}) with $y=\sgn V_0$,
and $p(V_0)$ is the Maxwell-Boltzmann distribution
\begin{align}
p(V_0)=\frac{1}{\sqrt{2\pi}V_T} \exp \left( -\frac{V_0^2}{2V_T^2} \right).
\end{align}
Here, $V_T = \sqrt{k_{\rm B}T/m}$ is the thermal velocity.

\begin{figure}[t]
\begin{center}
\includegraphics[scale=0.45]{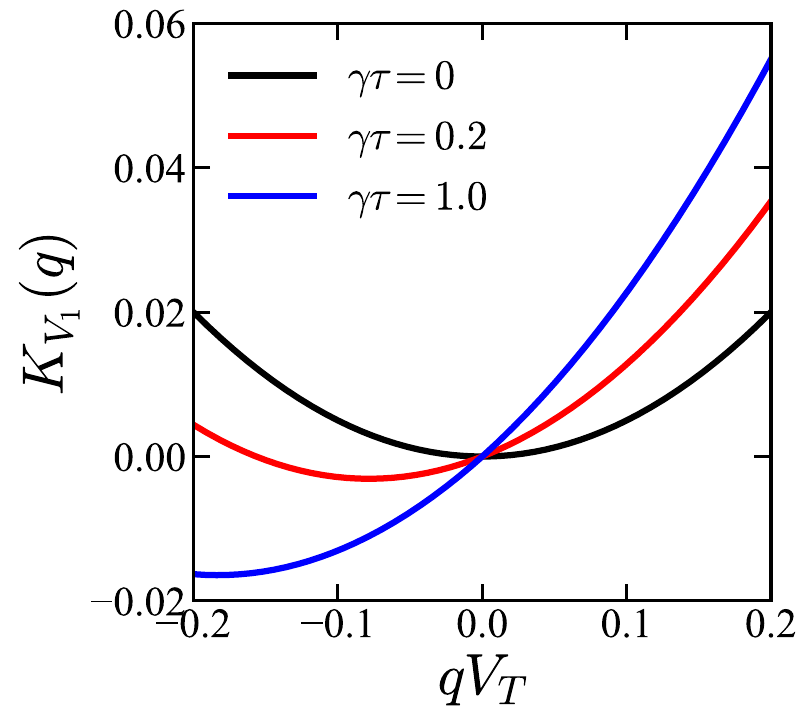}
\end{center}
\caption{
Cumulant generating function $K_{V_1}(q)$ for the single-measurement case, plotted as a function of 
$qV_T$ [see Eqs.~(\ref{Eq:Result-K}) and (\ref{Eq:Ipm})]. 
For $\gamma_\pm=\gamma(1\mp\delta)$ with $\delta=0.3$, we vary $\gamma\tau$ over $0$ (black), 
$0.2$ (red), and $1.0$ (blue), where $\tau$ denotes the measurement time.
All curves satisfy the normalization condition $K_{V_1}(0)=0$.
}
\label{Fig:K-result}
\end{figure}

We evaluate the integral over the initial velocity $V_0$.
Since the relaxation rate takes the value $\gamma_\pm =\zeta_\pm/m$ depending on the 
sign of $V_0$, the CGF can be written as $K_{V_1}(q) = \ln ( I_+ + I_- )$, where
\begin{align}
I_\pm & = \pm \exp\left[ q^2 V_T^2 e^{-\gamma_\pm \tau} \sinh(\gamma_\pm \tau) \right] 
\nonumber\\
& \quad \times \int_{0}^{\pm\infty} dV_0 \, \exp\left( q V_0 e^{-\gamma_\pm \tau} \right) p(V_0).
\nonumber \\
& = \frac{1}{2} \exp\left( \frac{q^2 V_T^2}{2} \right)
\left[ 1 \pm \erf \left( \frac{q V_T}{\sqrt{2}}  e^{-\gamma_\pm \tau} \right) \right].
\label{Eq:Ipm}
\end{align}
Hence, we obtain the CGF in Eq.~(\ref{Eq:Result-K}).
In Fig.~\ref{Fig:K-result}, the above CGF is plotted as a function of $qV_T$ for different
$\gamma \tau$ values, where $\gamma$ is defined by $\gamma_\pm = \gamma(1 \mp \delta)$ and 
the drag asymmetry parameter $\delta$ is chosen here as $\delta = 0.3$.

By taking the first and second derivatives of the CGF with respect to $q$, we obtain Eqs.~(\ref{Eq:Result-V1})
and (\ref{Eq:Result-V1second}), respectively.
By taking higher derivatives with respect to $q$, the third and fourth cumulants of $V_1$ can be obtained as 
\begin{align}
\frac{\langle V_1^3\rangle_\mathrm c}{V_T^3} & = \frac{2}{(2\pi)^{3/2}}
\nonumber \\ 
& \times [(e^{-\gamma_+\tau}-e^{-\gamma_-\tau})^3 -\pi(e^{-3\gamma_+\tau}-e^{-3\gamma_-\tau})],
\label{Eq:Result-VVV1} 
\end{align}
\begin{align}
\frac{\langle V_1^4\rangle_\mathrm c}{V_T^4} & = \frac{1}{2\pi^2} (e^{-\gamma_+\tau}-e^{-\gamma_-\tau})^2
\nonumber \\
& \times [-3(e^{-\gamma_+\tau}-e^{-\gamma_-\tau})^2
\nonumber \\
& +4\pi(e^{-2\gamma_+\tau}+e^{-\gamma_+\tau}e^{-\gamma_-\tau}+e^{-2\gamma_-\tau})].
\label{Eq:Result-VVVV1}
\end{align}

\section{Multiple measurements and the steady state}
\label{AppMulti}

To obtain the CGF of $V_{n+1}$, we use Eq.~(\ref{Eq:IOMVP0}) 
\begin{align}
K_{V_{n+1}}(q) = \ln \int_{-\infty}^{\infty} dV_n \, \exp\left[ \Omega_{V_{n+1}}^\ast(q, V_n) \right] p(V_n),
\end{align}
where $\Omega_{V_{n+1}}^\ast (q, V_n)$ is given by Eq.~(\ref{Eq:obtainedJx}) with $y=\sgn V_n$,
and $p(V_n)$ is the Gaussian distribution  
\begin{align}
p(V_n)=\frac{1}{\sqrt{2\pi \langle V_n^2\rangle_\mathrm c}} 
\exp\left[ -\frac{(V_n-\langle V_n\rangle_\mathrm c)^2}{2\langle V_n^2\rangle_\mathrm c} \right].
\label{steadyGaussian}
\end{align}

Since the relaxation rate takes the value $\gamma_\pm$ depending on the sign of $V_n$, the CGF can be written as
$K_{V_{n+1}}(q) = \ln (J_+ + J_- )$, where 
\begin{align}
J_\pm & = \pm \exp\left[ \frac{q^2 V_T^2}{2} \left( 1 - e^{-2\gamma_\pm\tau} \right) \right] \nonumber\\
& \quad \times \int_{0}^{\pm\infty} dV_n \, \exp (q V_n e^{-\gamma_\pm \tau} ) p(V_n)
\\
& = \frac{1}{2} \exp\left[ \frac{q^2 V_T^2}{2} \left( 1 - e^{-2\gamma_\pm\tau} \right) \right] \nonumber\\
& \quad \times \exp\left[ \frac{q e^{-2\gamma_\pm\tau}}{2} \left( q \langle V_n^2\rangle_{\mathrm{c}} + 2\langle V_n\rangle_{\mathrm{c}} e^{\gamma_\pm\tau} \right) \right] \nonumber\\
& \quad \times \left[ 1 \pm \erf \left( \frac{ q\langle V_n^2\rangle_{\mathrm{c}} + \langle V_n\rangle_{\mathrm{c}} e^{\gamma_\pm\tau} }{ \sqrt{2\langle V_n^2\rangle_{\mathrm{c}}} } e^{-\gamma_\pm\tau} \right) \right].
\end{align}

Expanding $K_{V_{n+1}}(q)$ in powers of $q$, we obtain the first and second cumulants:
\begin{align}
&\langle V_{n+1}\rangle_\mathrm c= \sqrt{\frac{\langle V_n^2\rangle_\mathrm c}{2\pi}}e^{-\langle V_n\rangle_\mathrm c^2/2\langle V_n^2\rangle_\mathrm c}(e^{-\gamma_+\tau}-e^{-\gamma_-\tau})\nonumber\\
&+\langle V_n\rangle_\mathrm c(e^{-\gamma_+\tau}+e^{-\gamma_-\tau})/2\nonumber\\
&+\langle V_n\rangle_\mathrm c(e^{-\gamma_+\tau}-e^{-\gamma_-\tau}) \erf(\langle V_n\rangle_c/\sqrt{2\langle V_n^2\rangle_c})/2,
\end{align}
\begin{align}
&\langle V_{n+1}^2\rangle_\mathrm c+\langle V_{n+1}\rangle_\mathrm c^2=\frac{V_T^2}{2}(2-e^{-2\gamma_-\tau}-e^{-2\gamma_+\tau})\nonumber\\
&+(\langle V_{n}^2\rangle_\mathrm c+\langle V_{n}\rangle_\mathrm c^2) (e^{-2\gamma_-\tau}+e^{-2\gamma_+\tau})/2\nonumber\\
&+\sqrt{\frac{\langle V_n^2\rangle_\mathrm c}{2\pi}}e^{-\langle V_n\rangle_\mathrm c^2/2\langle V_n^2\rangle_\mathrm c}\langle V_n\rangle_\mathrm c(e^{-2\gamma_+\tau}-e^{-2\gamma_-\tau})\nonumber\\
&+\frac{V_T^2}{2}(e^{-2\gamma_-\tau}-e^{-2\gamma_+\tau})\erf(\langle V_n\rangle_\mathrm c/\sqrt{2\langle V_n^2\rangle_\mathrm c})\nonumber\\
&+\langle V_n^2\rangle (e^{-2\gamma_+\tau}-e^{-2\gamma_-\tau})\erf(\langle V_n\rangle_\mathrm c/\sqrt{2\langle V_n^2\rangle_\mathrm c})/2.
\end{align}
These are the recurrence relations for $\langle V_n\rangle_{\mathrm{c}}$ and $ \langle V_n^2\rangle_{\mathrm{c}}$.
Under the assumption $\langle V_n\rangle_{\mathrm{c}}^2 \ll \langle V_n^2\rangle_{\mathrm{c}}$, the above recurrence relations 
can be simplified to 
\begin{align}
 \langle V_{n+1}\rangle_\mathrm c & \approx \sqrt{\frac{\langle V_n^2\rangle_\mathrm c}{2\pi}}(e^{-\gamma_+\tau}-e^{-\gamma_-\tau})
\nonumber \\
&\quad +\frac{\langle V_n\rangle_\mathrm c}{2}(e^{-\gamma_+\tau}+e^{-\gamma_-\tau}).
\label{Eq:recusionApp}
\\
\langle V_{n+1}^2\rangle_{\mathrm{c}} 
& \approx \frac{V_T^2}{2} (2 - e^{-2\gamma_{-}\tau} - e^{-2\gamma_{+}\tau} )
\nonumber\\
& \quad + \frac{\langle V_{n}^2\rangle_{\mathrm{c}}}{2} ( e^{-2\gamma_{-}\tau} + e^{-2\gamma_{+}\tau} ).
\label{RecursionSecond}
\end{align}

From Eq.~(\ref{RecursionSecond}), we find that $\langle V_\infty^2\rangle_{\mathrm{c}} =V_T^2$ in the steady state. 
Solving Eq.~(\ref{Eq:recusionApp}), we obtain the first cumulant 
$\langle V_{N}\rangle_{\mathrm{c}}$ in Eq.~(\ref{Eq:Result-VN}) for finite $N$.
If we assume $\langle V_0^2\rangle_{\mathrm{c}} = 0$, the second cumulant $\langle V_{N}^2\rangle_{\mathrm{c}}$
for finite $N$ becomes 
\begin{align}
\frac{\langle V_{N}^2\rangle_{\mathrm{c}}}{V_T^2} 
= 1 - \left( \frac{e^{-2\gamma_{+}\tau} + e^{-2\gamma_{-}\tau}}{2} \right)^{N}.
\end{align}


\end{document}